\documentclass[showpacs,twocolumn,aps,preprintnumbers,letterpaper]{revtex4}
\usepackage{amsmath,amssymb}
\usepackage{epsfig}
\usepackage{graphicx}
\usepackage{color}
\usepackage{amsmath}
\usepackage{slashed}
\usepackage{amsfonts}
\usepackage{epstopdf}
\graphicspath{{figures/}}
\usepackage[font=small,labelfont=bf]{caption}

\newcommand{\be}{\begin{equation}}
\newcommand{\ee}{\end{equation}}
\newcommand{\bear}{\begin{eqnarray}}
\newcommand{\eear}{\end{eqnarray}}
\newcommand{\ba}{\begin{array}}
\newcommand{\ea}{\end{array}}

\def\be{\begin{eqnarray}}
\def\ee{\end{eqnarray}}

\def\roughly#1{\mathrel{\raise.3ex\hbox{$#1$\kern-.75em%
\lower1ex\hbox{$\sim$}}}}

\def\barp{{\bar p}}

\begin{document}

\title{Chiral Random Matrix Model at Finite Chemical Potential:\\
Characteristic Determinant and Edge Universality}

\author{Yizhuang Liu}
\email{yizhuang.liu@stonybrook.edu}
\affiliation{Department of Physics and Astronomy, Stony Brook University, Stony Brook, New York 11794-3800, USA}

\author{ Maciej~A. Nowak}
\email{maciej.a.nowak@uj.edu.pl}
\affiliation{M. Smoluchowski Institute of Physics and Mark Kac Complex Systems Research Center, Jagiellonian University, PL-30348 Krakow, Poland}

\author{Ismail Zahed}
\email{ismail.zahed@stonybrook.edu}
\affiliation{Department of Physics and Astronomy, Stony Brook University, Stony Brook, New York 11794-3800 USA}


\date{\today}
\begin{abstract}
We derive an exact formula for  the stochastic evolution of the characteristic determinant of a class of deformed 
 Wishart matrices following from 
a chiral random matrix model of QCD at finite chemical potential.  In the WKB approximation, the characteristic determinant 
describes a sharp droplet of  eigenvalues  that deforms and expands at large stochastic times. Beyond the WKB limit, 
the edges of the droplet are fuzzy and  described by universal edge functions.  At the chiral point, the characteristic 
determinant in the microscopic limit is universal. Remarkably,  the physical 
chiral condensate at finite chemical potential may be  extracted
from current and quenched lattice Dirac spectra  using the universal edge scaling laws, without having to solve the QCD sign problem.
 \end{abstract}






\maketitle

\setcounter{footnote}{0}



%

\section{Introduction}

QCD breaks spontaneously chiral symmetry with a wealth of evidence 
in hadronic processes at low energies~\cite{BOOK}. First principle
lattice simulations strongly support that~\cite{LATTICE}.  The spontaneous
breaking is characterized by a large accumulation of  eigenvalues
of the Dirac operator near zero-virtuality~\cite{CASHER}. 
The zero virtuality regime is ergodic, and its neighborhood is diffusive~\cite{DISORDER}.

The ergodic regime of the QCD Dirac spectrum 
 is amenable to a chiral random matrix model~\cite{SHURYAK}.  In short, the model simplifies
the Dirac spectrum to its zero-mode-zone (ZMZ). The Dirac matrix is composed of hopping between 
$N$-zero modes and $N$-anti-zero modes because of chirality, which are  Gaussian sampled by the
maximum entropy principle.  The model was initially  suggested as a null dynamical
limit  of  the random instanton model~\cite{MACRO}.

QCD at finite chemical potential $\mu$ is subtle on the lattice due to the sign problem
\cite{SIGN}.  A number of effective  models have been proposed to describe the effects of matter in QCD with
light quarks~\cite{BOOK}. Chiral random matrix models offer a simple construct 
that retains some essentials of chiral symmetry both in vacuum and matter. For instance, 
in the chiral 1-matrix model finite $\mu$ is captured by a constant deformation of Gaussian matrix 
ensembles~\cite{STEPHANOV,US}. In the chiral 2-matrix model the deformation with $\mu$ is also
random~\cite{OSB,AKE}.  Chiral matrix models in matter were discussed by many~\cite{BLUE,RMANY}.
Recently both a universal shock analysis~\cite{NOWAK} and a hydrodynamical description of
the Dirac spectra were suggested~\cite{HYDROUS} both at zero and finite chemical potential.
 
The matrix models were shown to exhibit the same microscopic universality for small eigenvalues 
in the ergodic regime with vanishingly small $\mu^2$ in the large volume limit~\cite{RMANY}. The
chief observation is that in the weakly non-hermitean limit, the matrix models can be deformed in
a way that preserves the global aspects of the coset manifold under the general strictures of
spontaneously broken chiral symmetry and power counting in the so-called epsilon-regime~\cite{LEUT}.

At finite $\mu$ the distribution of  Dirac eigenvalues in the complex plane maps onto a
2-dimensional Coulomb gas whose effective action is mostly controlled by Coulomb's law, the
conformal and gravitational anomalies in 2-dimensions~\cite{HYDROUS}. These constraints
on the Dirac spectrum are beyond the range of chiral symmetry. The eigenvalues form Coulomb
droplets that stretch and break at finite $\mu$. The accumulation of the complex eigenvalues at the
edge of the droplet may signal a new form of universality unknown to chiral symmetry. The purpose
of this paper is to explore this possibility using the concept of characteristic determinants for a
unitary random matrix model at finite $\mu$.

With this in mind, we start by developing a stochastic evolution for a Wishart characteristic determinant 
associated to the standard chiral random matrix model for QCD Dirac spectra at finite chemical potential $\mu$, much along the lines suggested 
in~\cite{NOWAK} for the Ginibre ensemble. At finite $\mu$ the eigenvalues of the Dirac operator spread
in the complex plane. Their accumulation in droplets break spontaneously holomorphic
symmetry~\cite{STEPHANOV,US}. 
The characteristic determinant acts as an order parameter for this breaking being zero within the 
droplet and finite outside.  The evolution involves the eigenvalues as complex masses and their conjugates
and  is  diffusion-like asymptotically. The universal behavior of the characteristic
 determinant at the edge of the Ginibre droplet observed in~\cite{NOWAK} will be exploited here to derive
 a universal edge behavior for the  Dirac spectra at finite chemical potential.

 Finally, we note that  the study of deformed and non-hermitean Wishart matrices is interesting on its own as 
 it is of interest to many other areas such as telecommunications and finances, where issues of signal to noise 
 in the presence of attenuation or losses are relevant in designing more efficient routers or financial instruments
 \cite{NONHERM}.
 
 The main and new results of the paper are the following:
1/ The derivation of a closed evolution equation for the characteristic determinant 
for a non-hermitean deformation of Wishart matrices in relation to a 1-matrix model for 
the phase quenched QCD with  $N_f=4$ flavors at finite $\mu$; 2/ An explicit derivation 
of the envelope of the complex eigenvalues for the deformed Wishart matrices; 3/ An
explicit microscopic scaling law for the distribution of the deformed Wishart eigenvalues 
at the edge as traced by the envelope; 4/ An explicit scaling law on the real edge of the 
complex eigenvalue distribution that scales with the  chiral
condensate at finite $\mu$, allowing its extraction from current  and quenched Dirac spectra; 
5/ An  explicit microscopic scaling law for the  characteristic determinant at the 
chiral point that scales with infinitesimal $\mu$.

The organization of the paper is as follows:
In section II, we review the matrix model description of the partition function for $N_f$ flavors at finite $\mu$ and
its phase quenched approximation. In section III we show that a pertinent characteristic determinant is 
the phased quenched matrix-model partition function for $N_f=4$. We follow the recent work analysis in
\cite{NOWAK} and identify a mathematical time with a continuous deformation of the harmonic trap. We explicit
the evolution equation for the characteristic determinant and show that it is parabolic asymptotically. In section
IV we use the WKB method to solve the evolution equation for the boundary of the eigenvalue droplet in
leading order. In section V we derive an exact solution for the evolution of the characteristic determinant 
using the method of characteristics. In section VI we develop a semi-classical expansion 
of the exact solution to explicit the universal character of the edges of the droplet
of complex Dirac eigenvalues. At the chiral point, the characteristic determinant in the microscopic
limit follows from a universal Bessel kernel. Our conclusions are in section VII. In Appendix I we detail
an alternative scaling law for the characteristic determinant on the real edge of the complex spectrum. In Appendix II, we briefly quote the
results for the characteristic determinant following from a 2-matrix model and confirm its microscopic universality
at the chiral point.


%

\section{Chiral Matrix Model}

The  low lying eigenmodes of the QCD Dirac operator capture some aspects of the spontaneous
breaking of chiral symmetry both in vacuum and in matter.  Remarkably, their fluctuations  follow
by approximating the entries in the Dirac operator by purely random  matrix elements which 
are chiral (paired spectrum) and fixed by time-reversal symmetry (Dyson ensembles).
At finite $\mu$ the Dirac spectrum on the lattice is complex~\cite{BARBOUR,TILO}. The matrix
models at finite $\mu$~\cite{STEPHANOV,OSB} capture this aspect of the lattice
spectra and the nature of the chiral phase transition~\cite{BOOK,BLUE,RMANY}.

In this section, we will briefly review the salient features of the standard or 1-matrix 
model and explicit the relationship between the chiral Dirac ensemble 
and a deformed Wishart ensemble both at finite $\mu$. For that, 
consider the 1-matrix model at finite chemical potential for $N_f$ fundamental
quarks in the complex representation or $\beta=2$~\cite{STEPHANOV,US}

\begin{eqnarray}
\label{Z2}
&&{\bf Z}_{N_f}[\tau, {\bf z}=-im_f,\mu] =\left<{\rm det}\left({\bf z}-{\bf D}\right)^{N_f}\right>\nonumber\\
&&\equiv \int d{\bf T}\,d{\bf T}^\dagger\,{\bf P}(\tau, {\bf T})
\,\,{\rm det}\left( \begin{array}{cc}
{\bf z}& {\bf T}-i\mu  \\
{\bf T}^\dagger-i\mu & {\bf z}
\end{array} \right)^{N_f}
\end{eqnarray}
for equal quark masses $m_f$  in the complex representation. Here

\be
\label{M2}
{\bf P}(\tau, {\bf T})=e^{-\frac N\tau {\rm Tr}({\bf T}^\dagger {\bf T})}
\ee
and ${\bf T}$ is  ${(N+\nu)\times N}$ valued
complex matrix.
$\nu$ accounts for the difference between  the number of zero modes and anti-zero modes. 
The chiral Dirac matrix ${\bf D}$ in (\ref{Z2}) has $\nu$ unpaired zero modes and $N$ 
paired eigenvalues $\pm iz_j$  in the massless limit. The paired eigenvalues
delocalize and are represented by (\ref{Z2}).  The unpaired zero-modes decouple.
Throughout we will set $\nu=0$ and ${\bf T}$ is a square complex matrix.
In the vacuum, the Banks-Casher formula~\cite{CASHER} fixes the
dimensionful parameter to a constant $\tau\rightarrow 1/a$ with $\sqrt{a}=|q^\dagger q|_0/{\bf n}$ in terms of the 
massless quark condensate  and the density of zero modes ${\bf n}=N/V_4$.

\begin{figure}[h!]
 \begin{center}
 \includegraphics[width=7cm]{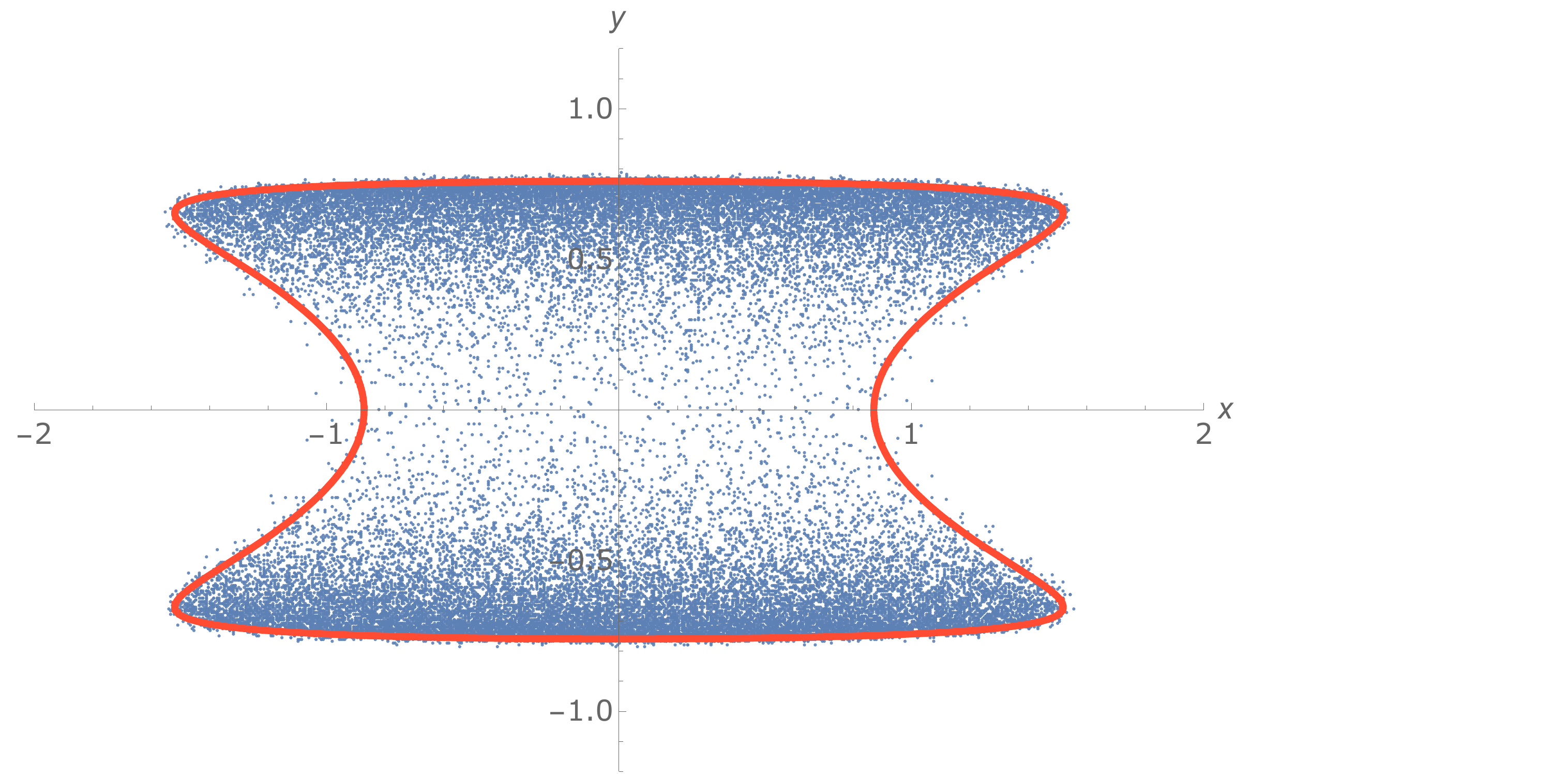}
  \caption{Eigenvalue distribution for the chiral Dirac matrices ${\bf D}$  for $\mu/\mu_c=0.9$ and $\tau=1$. }
    \label{fig_density1}
  \end{center}
\end{figure}

In Fig.~\ref{fig_density1} we display the distribution of eigenvalues following from the 1-matrix model 
with ${\bf T}$ sampled from a Gaussian ensemble of $200\times 200$ matrices with $\nu=0$ and 
$\mu=0.9$.  
The eigenvalue distribution forms
a connected droplet in the z-plane for $\mu<\mu_c=\sqrt{\tau}$, and splits into two  
droplets symmetric about the real-axis for $\mu>\mu_c=\sqrt{\tau}$,  restoring chiral symmetry~\cite{STEPHANOV,US}. 
In the spontaneously broken phase, all droplets are connected and symmetric about the real-axis. 
Some of these feature are shared by the lattice droplets of Dirac eigenvalues~\cite{BARBOUR,TILO}.

\begin{figure}[h!]
 \begin{center}
 \includegraphics[width=7cm]{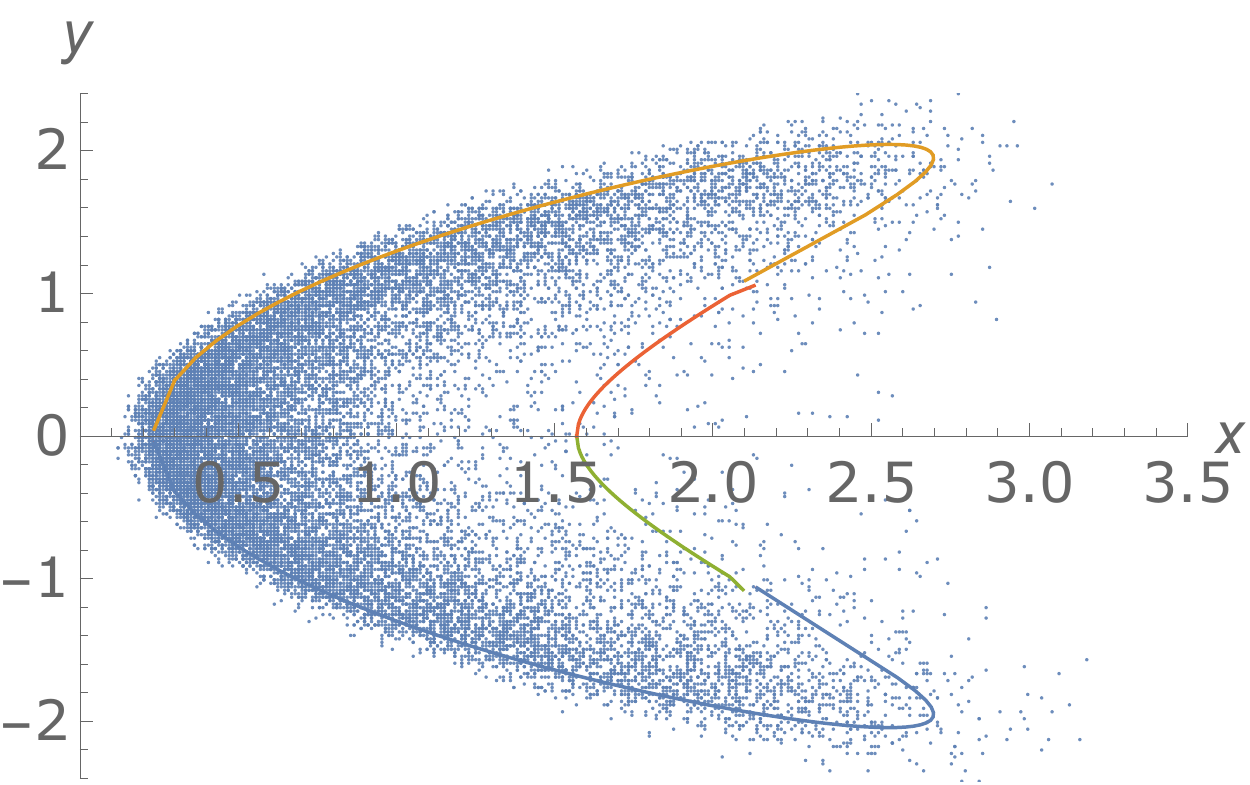}
  \caption{Eigenvalue distribution for the deformed Wishart ${\bf W}$ matrices for $\mu/\mu_c=0.9$  and $\tau=1$. }
    \label{fig_density2}
  \end{center}
\end{figure}

The complex nature of the eigenvalues entering in the determinant in (\ref{Z2}) yields to the so-called
sign problem when evaluating the complex partition function.  In lattice numerical analyses, the 
phase quenched 
partition function whereby the phase of the determinant is dropped is usually used. In the 1-matrix model
this amounts to using

\begin{eqnarray}
\label{ZQ}
{\mathbb Z}_{N_f}[\tau, {\bf z}=-im_f,\mu] =\left<{\rm det}\left|{\bf z}-{\bf D}\right|^{N_f}\right>
\end{eqnarray}
where the averaging is carried using (\ref{M2}).
In leading order in large $N$, the distribution of eigenvalues and its boundaries are the same for both
the unquenched and quenched partition functions since the phase factor is sub-leading in $1/N$. 
If we set $z={\bf z}^2+\mu^2$, (\ref{ZQ}) can be re-written as

\begin{eqnarray}
\label{ZQ2}
{\mathbb Z}_{N_f}[\tau, z, \mu] =
\left<\left|{\rm det}\left(z-{\bf W}\right)\right|^{\frac {N_f}2}\right>
\end{eqnarray}
with the deformed Wishart matrix

\be
{\bf W}={\bf T^\dagger}{\bf T}-i\mu({\bf T}^\dagger+{\bf T})
\label{D2}
\ee
The eigenvalue distribution for the deformed Wishart matrices (\ref{D2}) is shown in Fig.~\ref{fig_density2}
for $20\times 20$ matrices sampled from a similar
Gaussian ensemble with $\mu/\mu_c=0.9$. The droplet spreads and stretches vertically for increasing $\mu$ but
does not break. The density of eigenvalues within the droplets in Figs.~\ref{fig_density1},~\ref{fig_density2}  breaks spontaneously
conformal symmetry~\cite{STEPHANOV,US}. This breaking is best captured through the following
regulated partition function

\begin{eqnarray}
\label{ZQS}
{Z}_{N_f} [\tau, z, w,\mu]\equiv \left<
\left({\rm det}\left(\left|z-{\bf W}\right|^2+\overline{w}w\right)\right)^{\frac {N_f}4}\right>
\end{eqnarray}
which gives the partition function in the double limit 

\be
{\mathbb Z}_{N_f}[\tau, z,\mu] = \lim_{w\to 0}\lim_{N\to\infty}{Z}_{N_f} [\tau, z, w, \mu]
\ee

The measure in (\ref{Z2}-\ref{ZQS}) acts as a harmonic trap for the complex  eigenvalues that are 
deformed and split by the chemical potential. Following~\cite{NOWAK} we will identify 
$\tau$ with a mathematical and continuous  time deformation of the harmonic trap. (\ref{M2})
satisfies the formal matrix diffusion equation 

\be
\label{D1}
N\partial_\tau {\bf P}=\frac{\partial^2{\bf P}}{\partial {\bf T}^\dagger \partial {\bf T}}\qquad{\rm with} \qquad {\bf P}(0,  {\bf T})\approx \delta({\bf T})
\ee
(\ref{D1}) shows that the diffusive equation 
is purely kinetic with no potential or pressure like contribution. This is to be contrasted with the many-body
hydrodynamics expansion of the Dirac eigenvalues where both kinetic and pressure terms are identified in
the Eulerian flow in~\cite{HYDROUS}.

\section{Diffusion}

In this section we will define a pertinent characteristic determinant that will be used to 
analyze the nature and evolution of the complex eigenvalues of the deformed Wishart
ensemble. We will show that the evolution of the characteristic determinant obeys a
non-local equation that is diffusion-like asymptotically.

Indeed, a simple understanding of the accumulation and diffusion of 
the eigenvalues of the Dirac operator in the  complex plane follows  by identifying 
the phase quenched and regulated partition function (\ref{ZQS}) for
$N_f=4$ with the characteristic determinant

\be
{\bf\Psi}(\tau, z, w)\equiv Z_{N_f=4}[\tau, z, w, \mu]
\label{DET1}
\ee
(\ref{DET1}) defines a $2N$-degree polynomial which asymptotes $|z|^{2N}$~\cite{BLUE}.
The zeros of the characteristic determinant are the complex eigenvalues of the
deformed Wishart matrix ${\bf W}$ in (\ref{D2}). They 
are related to the eigenvalues of the Dirac operator ${\bf D}$ in the
complex 2-plane by recalling the mapping  $z={\bf z}^2+\mu^2$.  The macroscopic
density of complex and deformed Wishart eigenvalues is

\be
\rho_W(\tau, z)=\lim_{w\to 0}\lim_{N\to\infty}\frac 1{N\pi} \partial_{z\bar z}^2\,{\rm ln}{\bf \Psi}(\tau,z,w)
\label{GE}
\ee
with ${\rm ln}{\bf \Psi}/N$ acting as a Coulomb-like potential at large $N$.
The corresponding  eigenvalue density for the
chiral Dirac operator as a function of $\tau$, follows from (\ref{GE})  through

\be
\rho_D(\tau, {\bf z}) =2 |{\bf z}| \,\rho_W(\tau, {\bf z}^2+\mu^2)
\ee
The eigenvalues  condense in a droplet, 
with ${\bf \Psi}\approx 0$ inside and ${\bf \Psi}/|z|^{2N}\approx 1$ 
(order parameter), as the corresponding pressure ${\rm ln}{\bf \Psi}$ 
changes sign across the droplet boundary (phase change)~\cite{US,BLUE}.

Unwinding the determinant in ${\bf \Psi}$ through a Grassmannian quark $q$ and conjugate quark $Q$,
yields

\begin{eqnarray}
\label{G1}
&&{\bf \Psi}(\tau, z,w)\equiv \left<e^{{\bf F}+{\bf G}}\right>\nonumber\\
&&\equiv \int d{\bf T} d{\bf T}^\dagger dq\, dq^\dagger\,dQ\, dQ^\dagger \,\, {\bf P}(\tau, {\bf T}) \nonumber \\
&&\times \,\,e^{q^\dagger  (z-{\bf W})q+Q^\dagger (\bar z- {\bf W}^\dagger) Q -\bar w q^\dagger Q+w Q^\dagger q}
\end{eqnarray}
where we have defined 

\begin{eqnarray}
&&{\bf F}=q^\dagger  (z-{\bf W})q+Q^\dagger (\bar z- {\bf W}^\dagger) Q  \nonumber\\
&&{\bf G}=-\bar w q^\dagger Q+w Q^\dagger q
\end{eqnarray}
Note that the complex eigenvalues $z,\bar z$ act as complex masses   for the pair of quark  bilinears
$q^\dagger q$, $Q^\dagger Q$,  while $w,\bar w$ act as complex mixing masses for
the pair of mixed quark bilinears $Q^\dagger q$ and $q^\dagger Q$. The formers
preserve holomorphy, while the latters do not~\cite{BLUE}. 
(\ref{G1}) obeys the evolution equation

 \begin{eqnarray}
\label{LONGX}
&& N\partial_\tau{\bf \Psi}(\tau, z, w,\mu)=\nonumber\\
&& \left<\left(-N(q^\dagger q+Q^\dagger Q ) \right.\right.\nonumber\\
&&-q^\dagger qq^\dagger {\bf W} q-Q^\dagger QQ^\dagger {\bf W}^\dagger Q
 -q^\dagger Q Q^\dagger {\bf W} q -Q^\dagger qq^\dagger  {\bf W}^\dagger Q\nonumber\\
&&+2i\mu \,Q^\dagger qq^\dagger {\bf T} Q
-2i\mu \,q^\dagger QQ^\dagger {\bf T} q\nonumber\\
&&\left.\left. -2\mu^2 \,q^\dagger QQ^\dagger q+\mu^2\, (q^\dagger q+Q^\dagger Q)\right)
\,e^{{\bf F}+{\bf G}}\right>
 \end{eqnarray}
where we have used (\ref{D1}) . With the help of  the identity

\begin{eqnarray}
\label{ID}
&&\left<\left(+2i\mu \,Q^\dagger qq^\dagger {\bf T} Q
-2i\mu \,q^\dagger QQ^\dagger {\bf T} q\right)\,\,e^{{\bf F}+{\bf G}} \right>\nonumber\\
&&=-4\mu^2 \tau^2\frac{\partial_w\partial_{\bar w}(\partial_w\partial_{\bar w}+\partial_z \partial_{\bar z})}{(1-\tau\partial_z)(1-\tau\partial_{\bar z})+\tau^2\partial_w\partial_{\bar w}}\left<e^{{\bf F}+{\bf G}}\right>\nonumber\\
\end{eqnarray}
most of the terms on the right-hand-side of (\ref{LONGX}) can be turned either to 
ordinary z-derivatives of $e^{\bf F}$ or some Grassmannian derivative of $e^{\bf F}$ or $e^{\bf G}$. 
The final result is a closed  but non-local evolution of the characteristic determinant  (\ref{G1}), i.e.

\begin{eqnarray}
\label{CLOSED}
N\partial_{\tau} {\bf \Psi}=&&\Bigg(-2(\partial_z+\partial_{\bar z})\Bigg.\nonumber\\
&&-(z\partial_z^2
+\bar z\partial_{\bar z}^2-(z+\bar z) \partial_{\bar w w}^2)\nonumber \\
&&-(\partial_z+\partial_{\bar z})(w\partial_w+\bar w\partial_{\bar w})\nonumber\\
&&+2\mu^2 \partial^2_{\bar w w}+\mu^2(\partial_z^2+\partial_{\bar z}^2)\nonumber \\
&&\Bigg.-\left(\frac{2\mu\tau}{N}\right)^2\frac{\partial_w\partial_{\bar w}(\partial_w\partial_{\bar w}
+\partial_z \partial_{\bar z})}{\left|1-\frac{\tau}{N}\partial_z\right|^2
+\left|\frac{\tau}{N}\partial_w\right|^2}\Bigg){\bf \Psi}\nonumber \\
\end{eqnarray}
which is diffusive-like or parabolic at asymptotic times, subject to the initial condition

\be
\label{INIT0}
{\bf \Psi}(\tau=0, z, w)=(|z_0|^2+|w_0|^2)^N
\ee
(\ref{CLOSED}-\ref{INIT0}) is the first main result of this paper.

The stochastic evolution of ${\bf \Psi}$ 
involves the evolution in both the normal $z, \bar z$ and mixed $w, \bar w$ masses to
allow for the spontaneous breaking of chiral symmetry as well as the spontaneous 
breaking of holomorphy, respectively. The spontaneous breaking of chiral symmetry is signaled by the
accumulation of Dirac eigenvalues around zero or ${\bf z}=0$ ($z=\mu^2$), i.e. a non-vanishing
$\left<q^\dagger q\right>$. The spontaneous breaking
of holomorphy is signaled by the spreading of Dirac eigenvalues in the complex plane, i.e.
a non-vanishing $\left<|q^\dagger Q|^2\right>$~\cite{STEPHANOV,US,BLUE}.

\section{WKB approximation}

In this section we will provide a WKB analysis of the non-local and diffusion-like equation 
for the characteristic determinant derived in the previous section.  We will use it to derive
a polynomial and parametric equation for  the time-dependent envelope of the complex 
eigenvalues for the deformed Wishart ensemble, and by mapping for the standard but 
diffusing Dirac ensemble at finite $\mu$.

(\ref{CLOSED}) is a non-local Schroedinger-like evolution equation in Euclidean time. 
Some insights to this evolution can be obtained
using the WKB method in the large $N (= 1/\hbar)$ limit. For that we identify 

\be
\label{WKB}
{\bf \Psi}\approx e^{N{\mathbb S}}
\ee
in (\ref{CLOSED}) and define the conjugate momenta $p_\xi=\partial_\xi{\mathbb S}$ with $\xi=z, \bar z, r=\sqrt{w\bar w}$.  
Note that in leading $N$, the eigenvalue density for the deformed Wishart class in 
(\ref{D2}) is given by

\be
\rho_W(\tau, z)= \lim_{w\to 0}\frac 1\pi \partial_{\bar z}p_z
\label{RHOX}
\ee
which is non-holomorphic inside the droplet.

\subsection{Hamilton-Jacobi Equations}

Using the rescaling $N\tau\rightarrow \tau$, the
effective semi-classical action ${\mathbb S}$ obeys

\be
\partial_{\tau}{\mathbb S} +{\mathbb H}(\tau,\xi,p_\xi)=0
\label{DIFF2}
\ee
with  the $\tau$-dependent and non-local  effective Hamiltonian

\begin{eqnarray}
{\mathbb H}=&&+zp_z^2+\bar zp_{\bar z}^2-(z+\bar z)p_r^2/4\nonumber\\
&&+r(p_z+p_{\bar z})p_r-\mu^2(p_z^2+p_{\bar z}^2)-\frac{\mu^2p_r^2}{2}\nonumber \\
&&+\mu^2 \tau^2\frac{p_r^2(\frac{p_r^2}{4}+p_zp_{\bar z})}{(1-\tau p_z)(1-\tau p_{\bar z})+\tau^2\frac{p_r^2}{4}}
\end{eqnarray}
The initial condition is $\mathbb S(0)={\rm ln}(|z_0|^2+r_0^2)$.

 The semi-classical equations are the standard Hamilton-Jacobi equations,

\begin{eqnarray}
&&\frac{d p_z}{d\tau}=-\frac{\partial \mathbb H}{\partial z}=-p_z^2+p_r^2/4\nonumber\\
&&\frac{d p_{\bar z}}{d \tau }=-\frac{\partial \mathbb H}{\partial \bar z}=-p_{\bar z}^2 +p_r^2/4\nonumber \\
&&\frac{dp_r}{d\tau}=-\frac{\partial \mathbb H}{\partial r}=-p_r(p_z+p_{\bar z})
\end{eqnarray}
which are readily integrated

\begin{eqnarray}
&&p_x(\tau)=\frac{x_0+\tau}{(x_0+\tau)^2+y_0^2+r_0^2}\nonumber\\
&&p_y(\tau)=\frac{-y_0}{(x_0+\tau)^2+y_0^2+r_0^2}\nonumber\\
&&p_r(\tau)=\frac{2r_0}{(x_0+\tau)^2+y_0^2+r_0^2}
\end{eqnarray}
and

\begin{eqnarray}
\label{CC1}
&&\frac{dz}{d\tau}=\frac{\partial \mathbb H}{\partial p_z}\nonumber\\
&&\frac{dr}{d\tau}=\frac{\partial \mathbb H}{\partial p_r}
\end{eqnarray}
which are in general involved.

\subsection{Expanding droplet boundary}

The initialization of the characteristic determinant through (\ref{WKB}) at the droplet edge or $r_0=0$, 
allows for a simplification of (\ref{CC1}) at the edge. Indeed, (\ref{CC1}) for $z(\tau)$  at the edge, yields

\be
\label{ZT}
\frac{z(\tau)-\mu^2}{z_0-\mu^2}=\left(1+\frac{\tau}{z_0}\right)^2
\ee
while (\ref{CC1}) for small $r(\tau)$ gives

\begin{eqnarray}
\label{DETAIL}
&&\frac{dr}{d\tau}=f_1(\tau)r+r_0f_2(\tau)\\
&&f_1(\tau)=\frac{1}{z_0+\tau}+\frac{1}{\bar z_0 +\tau}\nonumber\\
&&f_2(\tau)=-\frac{4\mu^2\left(1-\frac{\tau^2}{|z_0|^2}\right)+\left((z_0-\mu^2)(1+\frac{\tau}{z_0})+{\rm c.c.}\right)}
{(x_0+\tau)^2+y_0^2}\nonumber
\end{eqnarray}
The formal solution of (\ref{DETAIL}) is

\be
\label{TRANS}
r(\tau )=r_0\left(1+\int_{0}^{\tau} d\tau^\prime f_2(\tau')e^{-\int_0^{\tau^\prime} d\tau'' f_1(\tau'')}\right)e^{\int_{0}^\tau d\tau^\prime f_1(\tau^\prime)}
\ee
The boundary of the eigenvalue droplet is set by the  condition

\be
\label{BOUND}
1+\int_{0}^{\tau}f_2(\tau')e^{-\int_0^{\tau^\prime}f_1(\tau'')d\tau''}d\tau^\prime=0
\ee
Inserting 

\be
e^{-\int_0^{\tau^\prime}f_1(\tau'')d\tau''}=\frac{|z_0|^2}{(x_0+\tau^\prime)^2+y_0^2}
\ee
in (\ref{BOUND}) and using

\begin{eqnarray}
&&\int_{0}^{\tau}d\tau^\prime \frac{x_0^2+y_0^2-\tau'^2}{|(x_0+\tau')^2+y_0^2|^2}=\frac{\tau}{(x_0+\tau)^2+y_0^2}\nonumber\\
&&\int_{0}^{\tau}d\tau^\prime \frac{(x_0+iy_0+\tau^\prime)^2}{|(x_0+\tau')^2+y_0^2|^2}=\frac{\tau}{\bar z_0(\bar z_0+\tau)}
\end{eqnarray}
yield the polynomial condition for $z_0=x_0+iy_0$

\be
\label{BOUND2}
\frac{4\mu^2 \tau}{(x_0+\tau)^2+y_0^2}+\frac{z_0-\mu^2}{z_0}\frac{\tau}{\bar z_0+\tau}+\frac{\bar z_0-\mu^2}{\bar z_0}\frac{\tau}{z_0+\tau}=1\nonumber \\
\ee

A simple check of the result (\ref{BOUND2}) follows for $\mu=0$,  for which  we have

\be
z_0+\bar z_0+2\tau=\frac{1}{\tau}(z_0+\tau)(\bar z_0+\tau)
\ee
The general solution is $z_0=\tau e^{i\theta}$. Inserting 
this solution in (\ref{ZT}),  we have

\be
z(\tau)=\tau(2+e^{-i\theta}+e^{i\theta})=4\,{\rm cos}^2(\theta/2)
\ee
which is the support of the $\tau$-expanding Wishart  line segment on the
real-axis, i.e. $[0,4\tau]$. For general $\mu$,  (\ref{BOUND2}) yields the 
expanding 4 branches (${\bf s, s}^\prime=\pm$)

\begin{eqnarray}
&&y_0^{{\bf ss}^\prime }(x_0)=\\
&&{\bf s}\tau \left(-\frac{x_0^2}{\tau^2}+\left(\frac 12 +\frac{\mu^2}{\tau}\right)
\left(1+{\bf s}^\prime\left(1-\frac{8\mu^2 x_0}{(\tau+2\mu^2)^2}\right)^{\frac 12}\right)\right)^{\frac 12}
\nonumber
\label{YX}
\end{eqnarray}
which once inserted in (\ref{BOUND2}) give a parametric description of the evolving 
$\tau$-expanding boundary,  as the envelope of the eigenvalues of the deformed
Wishart eigenvalues  as shown in Fig.~\ref{fig_density2}. The envelope for the 
distributions of Dirac eigenvalues in Fig.~\ref{fig_density1} follows from the deformed 
Wishart envelope by using  the inverse mapping ${\bf z}=\pm \sqrt{z-\mu^2}$. 
It is  in agreement with the original envelope obtained in~\cite{STEPHANOV,US}
using different arguments.

\section{Characteristic determinant}

In this section, we will provide a formal solution for the stochastically evolving
characteristic determinant for the deformed Wishart ensemble, that is exact for
finite size $N$ and time $\tau$. For that, we will  use 
the method of characteristics to solve exactly the partial differential equation 
(\ref{CLOSED}) in Fourier space.

A formal but exact solution to the diffusion-like equation (\ref{LONGX}) for the characteristic determinant
can be obtained by recasting (\ref{CLOSED})  in conjugate or Fourier space. Specifically,

\be
\label{FOURIER}
{\bf \Psi}(\tau, z, w)=\int \frac{d^2k}{(2\pi)^2}\frac{d^2p}{(2\pi)^2}e^{ik\cdot w+ip\cdot  z}\,\,\tilde{\bf \Psi}(\tau,k,p)
\ee
 Taking the Fourier transform of   (\ref{CLOSED}) yields 

\be
\label{HCONJ}
N\partial_\tau\tilde{\bf \Psi}=\tilde{\mathbb H}\,\tilde{\bf \Psi}
\ee
with the conjugate Hamiltonian

\begin{eqnarray}
\label{HF1}
&&\tilde{\mathbb H}=2p_1+\frac{\mu^2}{2}(k^2+p_1^2-p_2^2)\nonumber\\
&&-\frac{\mu^2\tau^2}{4N^2}\frac{k^2(p^2+k^2)}{(1-\frac{\tau p_1}{2N})^2
+\frac{\mu^2\tau^2}{N^2}\frac{p_2^2+k^2}{4}}\nonumber \\
&&+\frac{p_1^2-p_2^2-k^2}{2}\partial_{p_1}+p_1p_2\partial_{p_2}+p_1k\partial_k
\end{eqnarray}
after the shifts $k\rightarrow -ik$ and $p\rightarrow -i p$. Here
$k$ is the conjugate to $\sqrt{w\bar w}\equiv r$.

\subsection{Characteristic lines}

The evolution equation (\ref{HCONJ}) is now first order with rational coefficients. It can be solved
by the characteristic method exactly. The characteristic lines are

\begin{eqnarray}
\label{CRCR}
&&\frac{d\tau}{ds}=N\nonumber\\
&&\frac{dp_1}{ds}=-\frac{p_1^2-p_2^2+k^2}{2}\nonumber\\
&&\frac{dp_2}{ds}=-p_1p_2\nonumber\\
&&\frac{dk}{ds}=-p_1k
\end{eqnarray}
They are  readily solved

\begin{eqnarray}
\label{INVERTP}
&&p_1(s)=\frac{p_{10}+\frac{s}{2}(k_0^2+p_{10}^2+p_{20}^2)}{(1+\frac{s}{2}p_{10})^2+\frac{s^2}{4}(k_0^2+p_{20}^2)}\nonumber\\
&&p_2(s)=\frac{p_{20}}{(1+\frac{s}{2}p_{10})^2+\frac{s^2}{4}(k_0^2+p_{20}^2)}\nonumber\\
&&k(s)=\frac{k_{0}}{(1+\frac{s}{2}p_{10})^2+\frac{s^2}{4}(k_0^2+p_{20}^2)}
\end{eqnarray}
and inverted

\begin{eqnarray}
\label{INVERTX}
&&p_{10}=\frac{p_1-\frac{s}{2}(p_1^2+p_2^2+k^2)}{(1-\frac{s}{2}p_1)^2+\frac{s^2}{4}(p_2^2+k^2)}\nonumber\\
&&p_{20}=\frac{p_2}{(1-\frac{s}{2}p_1)^2+\frac{s^2}{4}(p_2^2+k^2)}\nonumber\\
&&k_0=\frac{k}{(1-\frac{s}{2}p_1)^2+\frac{s^2}{4}(p_2^2+k^2)}
\end{eqnarray}
We note the identity ($p_0^2=p_{10}^2+p_{20}^2$)

\be
\label{IDENT}
\frac{p^2+k^2}{(1-\frac{sp_1}{2})^2+s^2\frac{p_2^2+k^2}{4}}=p_0^2+k_0^2
\ee

\subsection{Exact determinant}

We are now set to evaluate the exact $\tau$-evolution of ${\bf\Psi}$. 
Inserting (\ref{CRCR}-\ref{IDENT}) into (\ref{HF1}) yield


\be
\label{INT}
\frac{d{\rm ln} \tilde{\bf \Psi}}{ds}=2p_1-\mu^2\left(-\frac{k^2+p_1^2-p_2^2}{2}+(p_0^2+k_0^2)\frac{s^2k^2}{4}\right)\nonumber\\
\ee
Using (\ref{INVERTP}) into (\ref{INT}) and undoing the derivative, we have

\begin{eqnarray}
\tilde{\bf \Psi}(\tau, k, p)=&&e^{-\mu^2\frac{s \left(-sp_{10} \left(p_0^2+k_0^2\right)-2 \left(p_{10}^2-p_{20}^2-k_0^2\right)\right)}{s^2 \left(k_0^2+p_0^2\right)+4 p_{10} s+4}}\nonumber\\
&&\times \frac{{\tilde{\bf \Psi}}_0(k_0,p_0)}{\left((1+\frac{sp_{10}}{2})^2+s^2\frac{p_{20}^2+k_0^2}{4}\right)^{-2}}
\end{eqnarray}

We now re-write $(k_0, p_0)$ in terms of $(k,p)$ and then undo the shifts through  $(k\rightarrow ik, p \rightarrow ip)$.
The results are

\begin{eqnarray}
\label{EXACT00}
&&\tilde{\bf \Psi}(\tau, k, p)=\mathbb K (\tau, k,p){\tilde{\bf\Psi}}_0(k_0(k,p),p_0(k,p))\nonumber\\
&&\mathbb K (\tau, k,p)=\frac{((2-p_1s)^2+s^2(p_2^2+k^2))^{-2}}{16}e^{-\mu^2{\bf S}(k,p,\tau)}\nonumber\\
&&{\bf S}(\tau, k,p)=\frac{s \left(sp_{1} \left(p^2+k^2\right)-2 \left(p_{1}^2-p_{2}^2-k^2\right)\right)}{s^2 \left(k^2+p^2\right)-4 p_{1} s+4}
\end{eqnarray}
In terms of the initial variables, the formal solution for $\mathbb K$  is

\be
\mathbb K (\tau, k_0,p_0)=\frac{e^{-\mu^2\frac{s \left(-sp_{10} \left(p_0^2+k_0^2\right)-2 \left(p_{10}^2-p_{20}^2-k_0^2\right)\right)}{s^2 \left(k_0^2+p_0^2\right)+4 p_{10} s+4}}}{\left((1+\frac{sp_{10}}{2})^2+s^2\frac{p_{20}^2+k_0^2}{4}\right)^{-2}}
\ee
The initial condition in Fourier space is

\be
{\tilde{\bf \Psi}}_0(k,p)=(\nabla_p^2+\nabla_k^2)^N \delta^2(k) \delta^2(p)
\ee
Thus 

\begin{eqnarray}
\label{EXACT1}
&&{\bf \Psi}(\tau,z,w)=\\
&&\left((\nabla_{p_0}^2+\nabla_{k_0}^2)^Ne^{p(p_0,k_0)\cdot z+k(k_0,p_0)\cdot w}{\mathbb J}\mathbb K (\tau, k_0,p_0)\right)_{p_0=k_0=0}
\nonumber
\end{eqnarray}
Here, ${\mathbb J}\equiv {\mathbb J}(k,p; k_0,p_0)$ is the Jacobian for the variable transformation 
$(k_0,p_0)\rightarrow( k,p)$ evaluated at $(k_0,p_0)$, i.e.

\be
{\mathbb J}\equiv 
\left(\left(1+\frac{\tau p_{10}}{2}\right)^2+\frac{\tau^2}4 ({p_{20}^2+k_0^2})\right)^{-3}\equiv\frac 1{F_0^{3}}
\ee
Although (\ref{EXACT1}) is exact for finite $N$, taking the large $N$ limit and assessing its corrections is in general more subtle.

\section{Universality at the Edge}

 The depletion of the zeros away from the droplet is captured by the way ${\bf \Psi}$ departs
 from zero away from the sharp boundary. The microscopic and universal changes in the eigenvalue density
 at the edges are commensurate with the microscopic rate of depletion of the zeros of the characteristic determinant. 
 We now develop a semi-classical expansion in $1/N$ and a pertinent microscopic re-scaling at the edge 
 to show this.

 \subsection{Saddle point approximation}
 
 To explicit this universal behavior, it is more appropriate to insert (\ref{EXACT00}) in (\ref{FOURIER}) 
and carry a semi-classical expansion around the saddle point  in terms of the initial coordinates
$(k_0, p_0)$. For notational convenience in this section we will re-label the coordinates
$(k_0, p_0)$  by $(k,p)$, and  the previous coordinates $(k, p)$ by $(K,P)$. 
This means that $P\equiv P(p,k)$ and $K\equiv K(p,k)$. With this in mind, we have

\begin{eqnarray}
\label{NEW}
&&{\bf \Psi}(\tau,z,w)=\\
&&\int \frac{d^2k}{(2\pi)^2}\frac{d^2p}{(2\pi)^2}d^2z^\prime 
d^2w^\prime\frac{e^{Nf(\tau,p,k,z,w,z^\prime, w^\prime)}}{\left((1+\frac{\tau p_1}{2})^2
+\tau^2\frac{p_2^2+k^2}{4}\right)}\nonumber
\end{eqnarray}
with 
\begin{eqnarray}
\label{fP}
&&f(\tau,p,k,z,w,z^\prime,w^\prime)=
P(p,k)\cdot z+K(p,k)\cdot w\nonumber\\
&&-p\cdot z^\prime-k\cdot w^\prime-\mu^2{\bf S}(p,k)+{\rm ln}(|z^\prime|^2+|w^\prime|^2)\\\nonumber\\
&&{\bf S}(p,k)=\frac{p_1+\frac{\tau}{2}(p_1^2+p_2^2-k^2)}{(1+\frac{\tau p_1}{2})^2+\tau^2\frac{p_2^2+k^2}{4}}-p_1
\end{eqnarray}

The saddle point corresponds to $\partial_{z^\prime,w^\prime, p,k}f=0$, which are respectively

\begin{eqnarray}
\label{SADDLEX}
&&\frac{\partial P}{\partial p}z+\frac{\partial K}{\partial p}w+\mu^2\frac{\partial S}{\partial p}=z^\prime\nonumber\\
&&\frac{\partial P}{\partial k}z+\frac{\partial K}{\partial k}w+\mu^2\frac{\partial S}{\partial k}=w^\prime\nonumber\\
&&p_i=\frac{2z^\prime_i}{|z^\prime|^2+|w^\prime|^2}\nonumber\\
&&k=\frac{2r^\prime}{|z^\prime|^2+|r^\prime|^2}
\end{eqnarray}
Near the boundary the first saddle point equation in (\ref{SADDLEX}) reduces to

\be
(z-\mu^2)\frac{z_0^2}{(\tau+z_0)^2}+\mu^2=z_0
\ee 
in agreement with (\ref{ZT}). The second saddle point equation in (\ref{SADDLEX}) becomes

\begin{eqnarray}
&&\lim_{r^\prime\to 0}\left(\frac{r^\prime}{k}\right)\frac{2(z_0+\tau)(\bar z_0+\tau)}{|z_0|^2}=\nonumber \\ 
&&\tau\left(4\mu^2+\frac{z_0-\mu^2}{z_0}(z_0+\tau)+\frac{\bar z_0-\mu^2}{z_0}(\bar z_0+\tau)\right)
\end{eqnarray}
which reduces to (\ref{BOUND2}) after using the last  two saddle point equations in (\ref{SADDLEX}).
In principle, the inversion of the above saddle point equations will determine $(z,w) $ as a function of $(z^\prime,w^\prime)$.
In practice, this inversion is involved. Fortunately, at the boundary there are simplifications since 
$(w=0, w^\prime=0)$, and since (\ref{BOUND2}) and (\ref{SADDLEX}) relate the saddle point initial positions to the current
positions.

\subsection{Microscopic correction}

The correction to the saddle point in momenta will be sought in holomorphic coordinates, i.e

\begin{eqnarray}
\label{PEX}
&&p_1-ip_2=p\nonumber\\
&&x_1p_2+x_2p_2=\frac{1}{2}( pz+\bar p \bar z)
\end{eqnarray}
and by expanding around the boundary using the following microscopic
rescalings

\begin{eqnarray}
\label{CEX}
&&z=z(z_0)+\frac{\delta z}{\sqrt{N}}\nonumber\\
&&z^\prime=z_0+\frac{\delta z^\prime}{\sqrt{N}}\nonumber\\
&&w^\prime=\frac{\delta r^\prime}{N^{\frac 14}}\nonumber\\
&&p=p_0(z_0)+\frac{\eta}{\sqrt{N}}\nonumber\\
&&k=\frac{\omega}{N^{\frac 14}}
\end{eqnarray}
The re-scaling in $\delta z^\prime, \delta z/\sqrt{N}$ at the boundary is natural, since the droplet area scales as ${\cal A}\approx N$
to keep the density of eigenvalues finite, while its length grows as ${\sqrt{{\cal  A}}}\approx \sqrt{N}$.  
Inserting (\ref{PEX}-\ref{CEX}) in (\ref{FOURIER}) and expanding to order $N^0$ at large $N$, we obtain

\begin{eqnarray}
&&N( f-f_0)-{\rm ln}\left(\left(1+\frac{\tau p_1}{2}\right)^2+\tau^2\frac{p_2^2+k^2}{4}\right)\approx \nonumber\\
&&{\bf Q}(\eta, \omega, \delta z,\delta z^\prime,\delta r^\prime)
+\sqrt{N}\left(\frac{\delta z}{z_0+\tau}+\frac{\delta \bar z}{\bar z_0+\tau}\right)\nonumber \\
&&+\sqrt{N}\left(\frac{(\delta r^\prime)^2}{|z_0|^2}+\frac{\omega^2|z_0|^2}{4}-\omega\cdot \delta r^\prime\right)\nonumber \\
&&-\frac{1}{2}\left(\frac{(\delta z^{\prime})^2}{z_0^2}+\frac{(\bar \delta z^{\prime})^2}{\bar z_0^2}\right)
- (\delta r^\prime)^2\left(\frac{\delta z^\prime}{z_0}+\frac{\delta \bar z^\prime}{\bar z_0}\right)-\frac{(\delta r^\prime )^4}{2|z_0|^4}\nonumber \\
\end{eqnarray}
with $f_0$ the value of $f$ in (\ref{NEW}) at the saddle point and

\begin{eqnarray}
\label{LONG}
&&{\bf Q}(\eta, \omega, \delta z,\delta z^\prime,\delta r^\prime)= \nonumber\\
&&-\frac{\tau z_0^3(z-\mu^2)}{4(z_0+\tau)^3}\eta^2-\frac{\tau \bar z_0^3(\bar z-\mu^2)}{4(\bar z_0+\tau)^3}\bar \eta^2\nonumber \\
&&+\left(\frac{z_0^2}{2(z_0+\tau)^2}\delta z -\frac{\delta z^\prime}{2}\right )\eta+\left(\frac{\bar z_0^2}{2(\bar z_0 +\tau)^2}\delta \bar z-\frac{\delta{\bar z}^\prime}{2}  \right)\bar \eta\nonumber \\&&+\frac{\tau |z_0|^2\omega^2}{4|(z_0+\tau)|^2}(\frac{z_0\delta z}{z_0+\tau}+\frac{\bar z_0\delta \bar z}{\bar z_0+\tau})-\frac{\tau^2\omega^4|z_0|^4}{16|(z_0+\tau)|^2}\nonumber \\
&&-\frac{\tau \omega^2|z_0|^2(\eta(\bar z_0+\tau)(z_0)+\bar \eta( z_0+\tau)\bar z_0}{8|(z_0+\tau)|^2}\nonumber \\&&-\frac{\tau^2\omega^2|z_0|^2}{8|z_0+\tau|^2}(\eta(z_0-\mu^2)+\bar \eta(\bar z_0-\mu^2))
\end{eqnarray}
Here $z=z(\tau)$ is defined in (\ref{ZT}).
Under the shift 

\be
\omega=\frac{2\delta r^\prime}{|z_0|^2}+\alpha
\ee
the $\alpha$-integration is subleading and decouples.
 With this in mind, we can  re-structure  and simplify ${\bf Q}$ in (\ref{LONG}) as

\begin{eqnarray}
&&{\bf Q}(\eta, \omega, \delta z,\delta z^\prime,\delta r^\prime)=\nonumber\\
&&Q(\eta)+\bar Q(\bar \eta  )+(\bar \eta,\eta)\cdot(\bar J,J)+Q_3(\delta r^\prime,\delta z)+Q_4(\delta  r^\prime)\nonumber \\
\end{eqnarray}
with

\begin{eqnarray}
&&Q(\eta)=-\frac{\tau z_0(z_0-\mu^2)}{4(z_0+\tau)}\eta^2\nonumber\\
&&J=\frac 12 \frac{z_0^2}{(z_0+\tau)^2}\delta z-\frac 1{2}{\delta z^\prime}-\frac{\tau (\delta r^{\prime})^2((1+\frac{\tau}{\bar z_0})+\tau\frac{z_0-\mu^2}{|z_0|^2})}{2|(z_0+\tau)|^2}\nonumber \\
&&Q_3=\frac{\tau (\delta r^{\prime})^2}{|z_0|^2|(z_0+\tau)|^2}
\left(\frac{z_0\delta z}{z_0+\tau}+\frac{\bar z_0\delta \bar z}{\bar z_0+\tau}\right)\nonumber \\
&&Q_4=-\frac{\tau^2(\delta r^\prime)^4}{|z_0|^4|z_0+\tau|^2}
\end{eqnarray}
The partial integration in (\ref{NEW}) of the quadratic contribution over $(\eta, \omega)$
gives

\begin{eqnarray}
\label{QUADX}
&&{\bf Q}(\delta z,\delta z^\prime,\delta r^\prime)=\nonumber\\
&&+\frac{J^2}{\frac{\tau z_0(z_0-\mu^2)}{(z_0+\tau)}}+\frac{\bar J^2}{\frac{\tau \bar z_0(\bar z_0-\mu^2)}{(\bar z_0+\tau)}}-\frac{\tau^2(\delta r^\prime)^4}{|z_0|^4|z_0+\tau|^2}
\nonumber \\&&+\frac{\tau (\delta r^{\prime})^2}{|z_0|^2|(z_0+\tau)|^2}
\left(\frac{z_0\delta z}{z_0+\tau}+\frac{\bar z_0\delta \bar z}{\bar z_0+\tau}\right)
\end{eqnarray}

\subsection{Microscopic egde profile}

The integration around $\delta z^\prime$ is Gaussian and is readily performed, leaving the last and non-Gaussian
integral in $\delta r^\prime$ undone. The result for the characteristic determinant close to the boundary
and in the microscopic limit is

\begin{eqnarray}
&&\Psi(\tau, z(z_0)+\delta z/\sqrt{N}, 0) \approx e^{Nf_0+\sqrt{N}(\frac{\delta z+\delta \bar z}{z_0+\tau})}\nonumber \\ 
&&\times \int_{0}^{\infty}xdx\,e^{-A(z_0)x^4+B(z_0,\delta z)x^2+C(z_0,\delta z)}
\label{SCA1}
\end{eqnarray}
with

\begin{eqnarray}
\label{PARA}
&&A(z_0)=\frac{1}{2|z_0|^4}+\frac{\tau^2}{|z_0|^4|z_0+\tau|^2}+\nonumber \\
&&\left({\left(\frac{1}{|z_0|^2z_0}-\frac{{\mathbb A}}{{\mathbb B}}\right)^2}{\left(\frac{1}{{\mathbb B}}-\frac{2}{z_0^2}\right)^{-1}}
-\frac{{\mathbb A}^2}{{\mathbb B}}+{\rm c.c.}\right)\nonumber\\
&&B(z_0,\delta z)=\delta z\left(\frac{\tau z_0}{|z_0|^2|z_0+\tau|^2(z_0+\tau)}\right.\nonumber\\
&&\left. -2{\mathbb C}\left({\frac{1}{z_0|z_0|^2}-\frac{2{\mathbb A}}{z_0^2}}\right)\left({1-\frac{2{\mathbb B}}{z_0^2}}\right)^{-1}\right)+{\rm c.c}\nonumber\\
&&C(z_0,\delta z)= \frac{z_0^2\delta z^2}{4\tau(z_0-\mu^2)(z_0+\tau)^2}\nonumber \\
&& \times \left(\frac{z_0}{z_0+\tau}-\frac{1}{\frac{z_0+\tau}{z_0}-\frac{2\tau(z_0-\mu^2)}{z_0^2}}\right)+ {\rm c.c.}
\end{eqnarray}
and

\begin{eqnarray}
&&{\mathbb A}=\frac{\tau((1+\frac{\tau}{\bar z_0})+\tau\frac{z_0-\mu^2}{|z_0|^2})}{2|z_0+\tau|^2}\nonumber\\
&&{\mathbb B}=\frac{\tau z_0(z_0-\mu^2)}{(z_0+\tau)}\nonumber\\
&&{\mathbb C}=\frac{z_0^2}{2(z_0+\tau)^2}
\end{eqnarray}
The depletion of the eigenvalues of the deformed Wishart matrices at the boundary as defined by
(\ref{ZT}) and (\ref{BOUND2}), follows the product of a Gaussian times an incomplete Error Function (Erfc) 

\begin{eqnarray}
\label{EDGEX}
&&\Psi(\tau, z(z_0)+\delta z/\sqrt{N}, 0) \approx e^{Nf_0+\sqrt{N}(\frac{\delta z+\delta \bar z}{z_0+\tau})}\nonumber\\
&&\times e^{\frac{B^2(z_0,\delta z)}{4A(z_0,\delta z)}+C(z_0,\delta z)}\int_{\frac{-B(z_0,\delta z)}{2A(z_0)}}^{\infty}dye^{-A(z_0)y^2}
\nonumber\\ &&\equiv  
e^{Nf_0+\sqrt{N}(\frac{\delta z+\delta \bar z}{z_0+\tau})}\nonumber\\
&&\times e^{\frac{B^2(z_0,\delta z)}{4A(z_0)}+C(z_0,\delta z)}\,{\rm Erfc}\left(\frac{-B(z_0,\delta z)}{2\sqrt{A(z_0)}}\right)
\end{eqnarray}
Recall that $z_0$ refers to the boundary value as a solution to (\ref{ZT}) and is valid throughout the edge
of the deformed Wishart droplet shown in Fig.~\ref{fig_density2}. We note that for $A>0$

\be
\lim_{B\to\ -\infty}\left(e^{\frac{B^2}{4A}}\,{\rm Erfc}\left(-\frac{B}{2\sqrt A}\right)\right)=0
\ee
(\ref{EDGEX}) is the second main result of this paper.

\subsection{Special edge points}

The depletion at the edge of the Wishart spectrum (\ref{EDGEX}) translates to a depletion at the edge of the Dirac
spectrum. We now make it explicit for the 4 cardinal points where the Dirac droplet crosses the eigenvalue
spectrum along the real and imaginary axes, e.g. see  Fig.~\ref{fig_density1}. Specifically, 
the edge of the Dirac droplet on the real axis corresponds to ${\bf y}(\tau)=0$. It maps onto
the Wishart boundary point $z_0=x_0+i0$ with $x_0$ in (\ref{EDGEX})  the real solution to 

\be
\label{X00}
x_0^4-2\tau\left(\frac {\tau}2 +\mu^2\right) x_0^2+2\mu^2\tau^2\,x_0=0
\ee
In general, the two real solutions to (\ref{X00}) are $x_0=\tau$ and $x_1/\tau=\frac{1}{2}(-1-\sqrt{1+8\mu^2/\tau})$.
They correspond to the outer and inner edge of the Wishart distribution in Fig.~\ref{fig_density2}.

The real solution $x_0=1$ yields  $x(\tau)=\mu^2+4(\tau-\mu^2)$ using (\ref{ZT}),  which is the outer edge along the real
axis in Fig.~\ref{fig_density2}.  It maps onto the two outer
edges along the real z axis of the Dirac spectrum in Fig.~\ref{fig_density1} or
${\bf z}=\pm 2\sqrt{\tau-\mu^2}$ using the 
Wishart to Dirac map $z=\mu^2+{\bf z}^2$. The 
corresponding edge parameters in (\ref{PARA}) are 

\begin{eqnarray}
&&A(x_0)= \frac{(\mu^2-\tau)^2}{16\mu^2\tau^5}\nonumber\\
&&B(x_0)=\frac{\mu^2-\tau}{16\mu^2\tau^3}(\delta z+\delta \bar z)\nonumber\\
&&C(x_0)=-\frac{1}{32\mu^2\tau}(\delta z^2+\delta \bar z^2)
\end{eqnarray}
and the scaling law (\ref{EDGEX}) on the Wishart envelope is ($\delta z=\delta x+i\delta y$) 

\begin{eqnarray}
&&\Psi(\tau, z(x_0)+\delta z/\sqrt{N}, 0) \approx \nonumber\\
&&e^{Nf_0}\,e^{\frac{\sqrt{N}\delta x}\tau}\,e^{\frac {\delta y^2}{16\mu^2\tau^2}}\,
{\rm Erfc}\left(\frac {\delta x}{4\mu\sqrt{{\tau}}}\right)
\label{EDGREAL}
\end{eqnarray}

Finally, the real solution $x_1/\tau=\frac{1}{2}(-1-\sqrt{1+8\mu^2/\tau})$ corresponds to the inner edge of the
Wishart distribution in Fig.~\ref{fig_density2} with $z(\tau)<\mu^2$. It maps onto the two outer edges along
the imaginary axis of the Dirac spectrum shown in Fig.~\ref{fig_density1}.
The  corresponding edge parameters in (\ref{PARA}) are too lengthy to report here.

\subsection{Application to $\mu=\frac{\sqrt{\tau}}2$}

To be more specific consider the special case of
$\mu/\sqrt{\tau}=1/2$,  for which the two real solutions are $z_0=x_0=\tau$ and $z_0=x_0=-(\sqrt{3}+1)\tau/2$.
The first solution corresponds to the outer edge along the real axis of the Wishart spectrum and
maps onto the outer edge of the Dirac spectrum also along the real axis. It gives

\begin{eqnarray}
\label{SDD}
&&A(z_0=\tau)=\frac{0.14}{\tau^4}\nonumber\\
&&B(z_0=\tau)=-\frac{0.19}{\tau^3}(\delta z+\delta \bar z)\nonumber\\
&&C(z_0=\tau)=-\frac{1}{8\tau^2}(\delta z^2+\delta \bar z^2)
\end{eqnarray}
The scaling law at the edge of the Wishart spectrum along the real axis is

\be
\Psi\approx e^{Nf_0}\, e^{\sqrt{N}\frac {\delta x}{\tau}}\,e^{\frac{\delta y^2}{4\tau^2}}\,{\rm Erfc}\left(\frac {0.5\,\delta x}\tau\right)
\ee
The second solution corresponds to the inner edge along the real axis of the Wishart spectrum ($z<\mu^2$)
and maps onto the outer edge of the Dirac spectrum along the imaginary axis. It gives

\begin{eqnarray}
&&A(z_0=-(\sqrt{3}+1)\tau/2)=\frac 3{\tau^4}\nonumber\\
&&B(z_0=-(\sqrt{3}+1)\tau/2)=\frac{11.19}{\tau^3}(\delta z+\delta \bar z)\nonumber\\
&&C(z_0=-(\sqrt{3}+1)\tau/2)=-\frac{6.96}{\tau^2}(\delta z^2+\delta \bar z^2)
\end{eqnarray}
with  $A>0$ in this case. Inserting the parameters in (\ref{EDGEX})  we have

\be
\Psi\approx e^{Nf_0}\,e^{\sqrt{N}\frac{\delta x}\tau}\,e^{27.83\frac{\delta x^2}{\tau^2}+13.92\frac{\delta y^2}{\tau^2}}\,
{\rm Erfc}\left(-\frac{6.46\,\delta x}{\tau}\right)
\ee

\subsection{Translation to Dirac}

The general result (\ref{EDGEX}) holds around the envelope or boundary of the deformed Wishart ensemble (\ref{D2})
as illustrated in Fig.~\ref{fig_density2}. 
Its translation to the  envelope of the Dirac ensemble 
as illustrated in Fig.~\ref{fig_density1},  follows from the mapping between the complex eigenvalues or
$z=\mu^2+{\bf z}^2$. An infinitesimal displacement $\delta z$ on the Wishart boundary or $z=z_0+\delta z/\sqrt{N}$, 
translates to the infinitesimal displacement $\delta {\bf z}$ on the Dirac boundary 

\begin{eqnarray}
\label{EXPX}
{\bf z}=&&{\bf z}_0+\frac{\delta {\bf z}}{\sqrt{N}}\nonumber\\
=&&\pm \left(z_0-\mu^2+\frac {\delta z}{\sqrt N}\right)^{\frac 12}\nonumber\\
\approx &&\pm \left({\bf z}_0
+\frac{\delta z}{2(z_0-\mu^2)^{\frac 12}\sqrt{N}}\right)
\end{eqnarray}
Therefore, (\ref{EDGEX}) maps onto the Dirac boundary through the substitution 

\be
\delta z\rightarrow \pm 2\sqrt{z_0-\mu^2}\,\delta{\bf z}
\label{TRANX}
\ee

In this spirit, the translation of the Wishart result (\ref{EDGREAL}) to Dirac follows  using the substitution (\ref{TRANX}),
with $z_0=\tau$ in this case. 
The scaling law at the real edge of the 
Dirac spectrum is sensitive to the chiral condensate, which follows from the large $N$
saddle point of the full or unquenched partition function (\ref{Z2}), 

\begin{eqnarray}
\label{QQX}
\left<\bar q  q\right>=&&\lim_{ m_f\to\ 0}\lim_{N\to\ \infty}
\left(-\frac 1{NN_f} \frac{\partial \,{\rm ln}\,{\bf Z}_{N_F}}{\partial i m_f}\right)\nonumber\\
=&&-2\,\tau\,\left(\tau-\mu^2\right)^{\frac 12}
\end{eqnarray}
which is seen to vanish for $\mu=\mu_c=\sqrt{\tau}$ in the massless case (second order transition).
This is remarkable, as it allows for a determination of the physical chiral condensate (\ref{QQX}) 
from the microscopic scaling law at the edge of the quenched Dirac spectrum. Indeed, 
if we set $\Sigma\equiv\left|\left<\bar q  q\right>\right|$ at finite $\mu$, the Wishart edge scaling law 
 (\ref{EDGREAL})  translates to the Dirac edge scaling law

\begin{eqnarray}
&&\Psi_D(\tau, {\bf z}_0+\delta {\bf z}/\sqrt{N}, 0) \approx \nonumber\\
&&e^{Nf_0}\,e^{\sqrt{N}\Sigma\frac{\delta {\bf x}}{\tau^2}}\,e^{\Sigma^2\frac {\delta {\bf y}^2}{16\mu^2\tau^3}}\,
{\rm Erfc}\left(\frac\Sigma{4\mu \tau}\frac {\delta {\bf x}}{\sqrt\tau}\right)
\label{EDGREALDIRAC}
\end{eqnarray}
While the exponent in the second factor in (\ref{EDGREALDIRAC}) grows initially with $\delta {\bf x}$,
it is countered by the rapid fall off of the complementary error function along the real axis. (\ref{EDGREALDIRAC})
is vanishingly small for positively large $\delta{\bf x}$. 

In~\cite{AKEX} it was shown that for a class of normal matrices, the spectral density is related  to 
the characteristic determinant for $w\rightarrow 0$, with a proportionality factor related to some pertinent
weight factor. This result suggests that in our case which is non-normal, the first two factors
in (\ref{EDGREALDIRAC}) may be part of  an underlying weight factor as they follow from a standard  $1/N$ 
saddle point approximation, i.e. order $N$ and order $N^0$.  The last factor in (\ref{EDGREALDIRAC}) does not. It emerges from
the specific $1/\sqrt{N}$ level spacing  law in (\ref{CEX}). We identify it with  the edge scaling law for 
the Dirac spectral density

\be
\rho_D(\tau, {\bf x}_0+\delta {\bf x}/\sqrt{N}, 0)\approx \frac 1{2\pi \tau}\,
{\rm Erfc}\left(\frac\Sigma{4\mu \tau}\frac {\delta {\bf x}}{\sqrt\tau}\right)
\label{DENSMICRO}
\ee
In the microscopic limit, the Dirac eigenvalue density along the real-axis follows the universal profile
of a complementary error function that is sensitive to the physical chiral condensate $\Sigma$ at finite $\mu$.
(\ref{DENSMICRO})  suggests a complementary scaling law for extracting $\Sigma$ from the Dirac spectrum.

\subsection{Check  at  the edge point $x_0=\tau$}

As a way to check on the general result (\ref{EDGEX}) we will re-analyze (\ref{NEW}) by 
trading $(P,K)\leftrightarrow (p,k)$. Using the complex coordinates (\ref{PEX}), this amounts to re-writing
(\ref{NEW}) as

\be
\label{NEWx}
\Psi(\tau, z, w)=\int \frac{d^2pd^2kd^2z^{\prime}d^2w^{\prime}}{(2\pi)^4}\,{\mathbb J}\, e^{Nf}
\ee
with

\begin{eqnarray}
\label{NEWxx}
&&f=\frac{1}{2}pz+\frac{1}{2}\bar p \bar z+k\cdot r\nonumber \\
&&-P(p,k)z^{\prime}-\bar P(p,k)\bar z^{\prime}-K(p,k)\cdot r^{\prime}+\frac{\mu^2}{2}(S+\bar S)\nonumber\\ 
&&+{\rm ln}(|z^{\prime}|^2+(r^{\prime})^2)
\end{eqnarray}
and 

\begin{eqnarray}
\label{NEWxxx}
&&P(p,k)=\frac{p(1-\frac{\tau \bar p}{2})-\frac{\tau k^2}{2}}{(1-\frac{\tau p}{2})(1-\frac{\tau \barp}{2})+\frac{\tau^2k^2}{4}}\nonumber\\
&&K(p,k)=\frac{k}{(1-\frac{\tau p}{2})(1-\frac{\tau \barp}{2})+\frac{\tau^2k^2}{4}}\nonumber \\
&&S(p,k)=\frac{p(1-\frac{\tau \bar p}{2})+\frac{\tau k^2}{2}}{(1-\frac{\tau p}{2})(1-\frac{\tau \barp}{2})+\frac{\tau^2k^2}{4}}-p
\end{eqnarray}
Here $1/{\mathbb J}={((1-\frac{\tau p}{2})(1-\frac{\tau \bar p}{2})+\frac{\tau^2k^2}{4})^2}$.

Around $x_0=\tau$  at the boundary, we will use the following microscopic rescaling

\begin{eqnarray}
\label{SCASCA}
&&z=4\tau-3\mu^2+\frac{\eta}{\sqrt{N}},p=\tau+\frac{p}{\sqrt{N}},k=\frac{k}{N^{1/4}}\nonumber \\
&&z^{\prime}=\tau+\frac{\delta z^{\prime}}{\sqrt{N}},r^{\prime}=\frac{r^{\prime}}{N^{1/4}}
\end{eqnarray}
and keep only terms that survive at large $N$. The result in leading order is

\be
N(f-f_0)\approx \frac{\sqrt{N}}{2}(p\eta+\bar p\bar \eta)+{\mathbb F}
\ee
with
\begin{eqnarray}
&&{\mathbb F}=-2p\delta z^{\prime}-2\bar p\delta z^{\prime}-\tau(\tau-\mu^2)(p^2+\bar p^2)\nonumber \\ 
&&+2\tau k^2(\delta z^{\prime}+\delta \bar z^{\prime})+4\tau^3k^2(p+\bar p)\nonumber\\
&&+2\tau^2k^2(\tau-\mu^2)(p+\bar p)
-4\tau^4k^4-4\tau k(p+\bar p)r^{\prime}\nonumber \\
&&-4\sqrt{N}kr^{\prime}+\sqrt{N}(r^{\prime}/\tau)^{2}+4\sqrt{N}\tau^2k^2\nonumber\\
&&-\frac{(\delta z^{\prime})^2+(\delta \bar z^{\prime})^2}{2\tau^2}-\frac{(r^{\prime})^4}{2\tau^4}
-\frac{(r^{\prime})^2}{\tau^3}(\delta z^{\prime}+\delta \bar z^{\prime})
\end{eqnarray}
Using  the shift $r^{\prime}=2k\tau^2+\frac{\alpha}{N^{1/4}}$, we can convert the $r^\prime$-integral to
an $\alpha$-integral which is Gaussian and decouples. The $\delta z^{\prime}$ integral can be 
undone. The result is

\begin{eqnarray}
\label{XNEWX}
&&\Psi(\tau, 4\tau-3\mu^2+\eta/\sqrt{N}, 0)\approx\nonumber\\
&& e^{Nf_0}e^{{\sqrt{N}\frac{(\eta+\bar \eta)}{2\tau}}}\int \frac{d^2pd^2k}{(2\pi)^4}e^{\frac{p\eta+\bar p\bar \eta}{2}+k\cdot r+{\mathbb G}(k,p)}
\end{eqnarray}
with

\be
{\mathbb G}(k, p)=2\tau\mu^2(p^2+\bar p^2)+2\tau^2k^2(\tau-\mu^2)(p+\bar p)
\ee
For $r=0$, the momentum integral in (\ref{XNEWX}) gives

\begin{eqnarray}
\label{COMPX}
&&\int _{0}^{\infty}kdk\,\left|e^{-\frac{1}{8\mu^2}(\frac{\eta}{2}+2\tau^2k^2(\tau-\mu^2))^2}\right|^2\nonumber\\
&&=\frac{1}{64\pi^2\mu^2(\tau-\mu^2)}e^{\frac{\delta y^2}{16\tau\mu^2}}\,{\rm Erfc}\left(\frac{x}{4\sqrt{\tau}|\mu|}\right)
\end{eqnarray}
Thus the scaling law for the characteristic determinant at the edge point $x_0=\tau$ and fixed but un-scaled $\mu$,  is

\be
\label{CHAMU}
&&\Psi(\tau, 4\tau-3\mu^2+\delta z/\sqrt{N}, 0)\approx\nonumber\\
&&\frac{ \tau^2e^{Nf_0}\,e^{{\sqrt{N}\frac{\delta x}{\tau}}}}{64\pi^2\mu^2(\tau-\mu^2)}\,e^{\frac{\delta y^2}{16\tau\mu^2}}\,
{\rm Erfc}\left(\frac{\delta x}{4\sqrt{\tau}\mu}\right)
\ee
for the Wishart ensemble and in agreement with (\ref{EDGREAL}). The translation to the real edge of the  Dirac ensemble
follows from the  substitution (\ref{TRANX}). The microscopic scaling law of the characteristic 
determinant near the real edge of the  complex Dirac spectrum (boundary of the zero mode zone) allows  for 
a reading of the quenched chiral condensate (\ref{QQX}) by fitting to the universal scaling function (\ref{CHAMU}) or its most
general form (\ref{ANXXX}) in Appendix I. This the third main result of this paper.

\subsection{Airy universality  at $\mu=0$}

For $\mu\rightarrow 0$  the result (\ref{CHAMU}) is singular. This feature is valid throughout
the envelope of the Wishart and Dirac spectra. The large $N$ limit and the $\mu=0$ do not
commute. Indeed, for $\mu=0$ the spectra are now real, and the new microscopic scaling laws 

\be
z=4\tau+\frac{\eta}{N^{\frac 23}},\qquad w=\frac{\omega}{N^{\frac 32}}
\ee
should replace (\ref{SCASCA}), with the new and re-scaled ansatz 

\begin{eqnarray}
&&\Psi(\tau, 4\tau+\eta/N^{\frac 23}, \omega/N^{\frac 32}) \approx \nonumber\\
&&\tau^{2N}e^{{N}^{\frac 13}\frac{(\eta+\bar\eta)}{2\tau}}\,\psi(\tau, \eta,\bar \eta, \omega)
\end{eqnarray}
To order $N^2$ and $N^{\frac 53}$, the equation (\ref{CLOSED}) is satisfied identically, irrespective of $\psi$. At order $N^{\frac 43}$, the 
new equation  fixes $\psi$

\be
\label{NRWYY}
-4\tau(\partial_{\eta}^2+\partial_{\bar \eta}^2)\psi+\frac{1}{4\tau^2}(\eta+\bar \eta )\psi+8\tau\partial_{\bar \omega \omega}^2\psi=0
\ee
The general solution to (\ref{NRWYY}) is

\be
\int d\lambda\,\kappa (\lambda)\, 
I_0\left(\sqrt{\frac{\lambda}{8\tau^2}}|\,\omega|\right)
\left|{\rm Ai}\left(2^{\frac 23}\left(\frac{\eta}{4\tau}-{\lambda}\right)\right)\right|^2
\ee
with $\kappa (\lambda)$ a general positive weight.
The microscopic determinant at the right edge of the Wishart ensemble
involves an Airy kernel. As expected, the mapping through $z={\bf z}^2$ at $\mu=0$ yields  an
Airy kernel for both edges of the real Dirac spectrum.

\section{Chiral Universality}

The mapping $z={\bf z}^2+\mu^2$ between the deformed Wishart ($z$) and Dirac $({\bf z})$ eigenvalues, shows that the zero
virtuality point in the Dirac spectrum at finite $\mu$ at ${\bf z}=0$,  maps onto the  $z=\mu^2$ point in the deformed
Wishart spectrum. For $\mu/\mu_c<1$, 
this point lies within the Dirac and Wishart droplets,  and moves out of both droplets for $\mu/\mu_c>1$. 
 We now analyze the nature of the  accumulation of  eigenvalues  around this point 
using the characteristic determinant.

\begin{figure}[h!]
 \begin{center}
 \includegraphics[width=7cm]{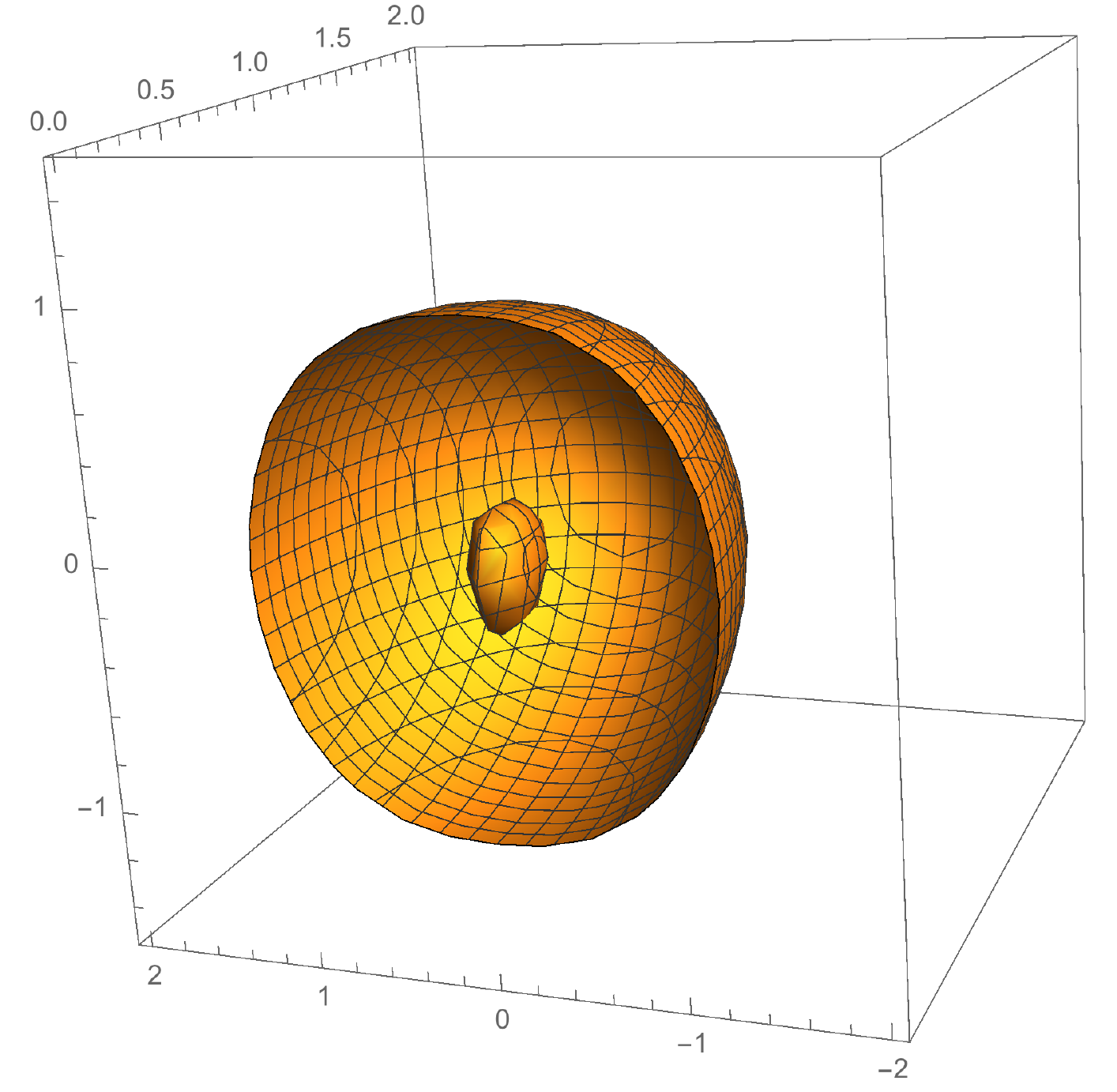}
  \caption{Saddle point surface $r_0(x_0, y_0)$ viewed along $0<x_0<2$ for $\mu^2=\frac 14$ and $\tau=1$. 
  The lateral axis is $y_0$, the height is $r_0$ and the depth is $x_0$. }
    \label{SF-POSITIVEx}
  \end{center}
\end{figure}

\begin{figure}[h!]
 \begin{center}
 \includegraphics[width=7cm]{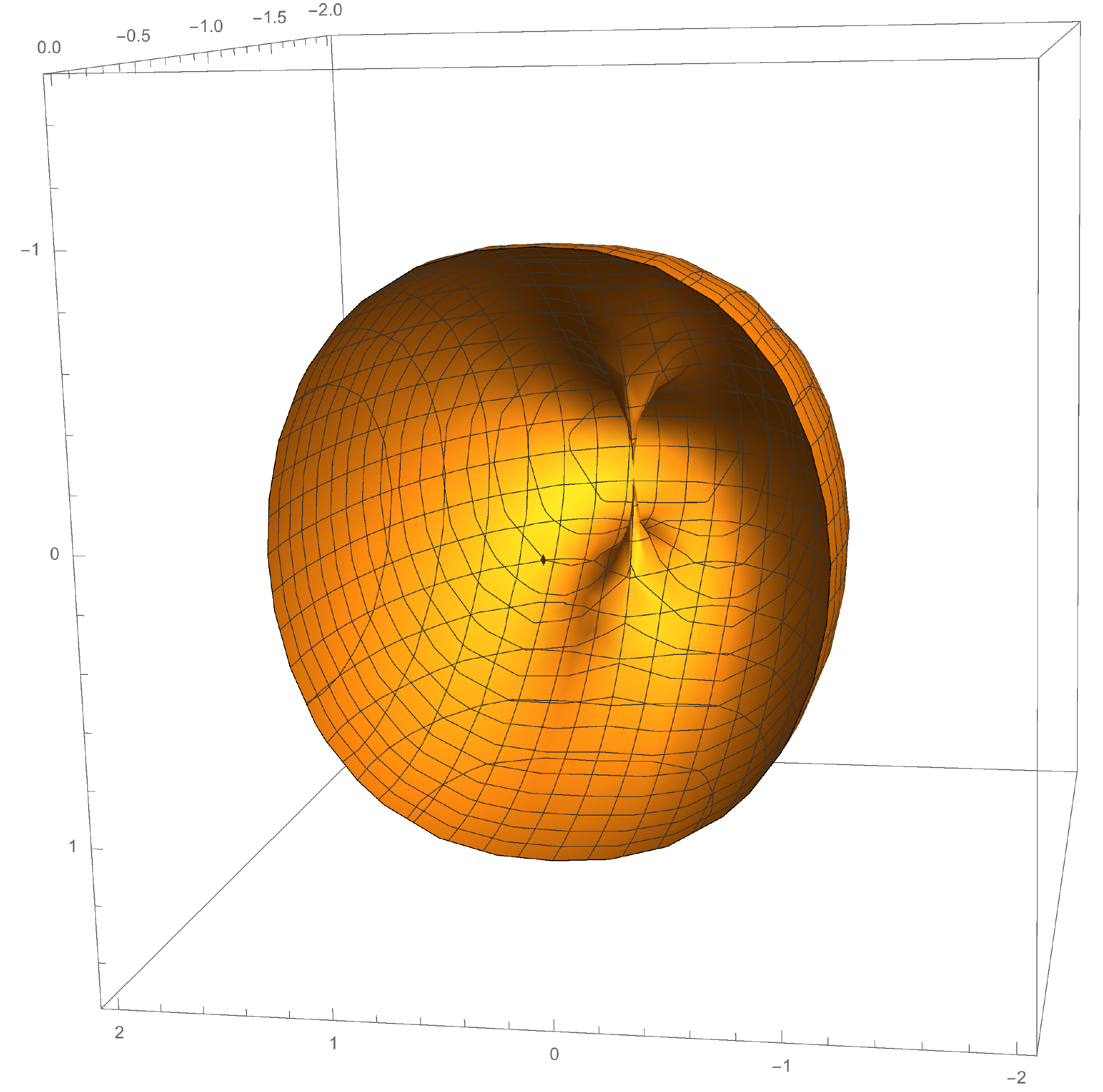}
  \caption{Saddle point surface $r_0(x_0, y_0)$ viewed along $-2<x_0<0$ for $\mu^2=\frac 14$ and $\tau=1$. 
 The lateral axis is $y_0$, the height is
  $r_0$ and the depth is $x_0$.  The pinch at $r_0(x_0, y_0)\equiv 0(-1, 0)$ is the chiral point in the deformed Wishart spectrum. }
    \label{SF-NEGATIVEx}
  \end{center}
\end{figure}

\subsection{Pinch at zero virtuality}

To analyze the saddle point equations (\ref{SADDLEX})  in the vicinity of $z\approx \mu^2$ for arbitrary $\mu^2$ and $w$, 
we will specialize to the case $\mu^2=1/4$ and $\tau=1$ for simplicity. Throughout this section $\tau=1$.
With this in mind, the second equation in (\ref{SADDLEX})
reads

\be
\label{LONGG}
(2x_0-2x_0^2-10x_0^3-2x_0^4+8x_0^5+4x_0^6-6y_0^2-10x_0y_0^2\nonumber \\-4x_0^2y_0^2+16x_0^3y_0^2+12x_0^4y_0^2-2y_0^4+8x_0y_0^4+12x_0^2y_0^4\nonumber \\4y_0^6-2r_0^2-10x_0r_0^2-4x_0^2r_0^2+16x_0^3r_0^2+12x_0^4r_0^2\nonumber \\-4y_0^2r_0^2+16x_0y_9^2r_0^2+24x_0^2y_0^2r_0^2+12y_0^4r_0^2-2r_0^4\nonumber \\+8x_0r_0^4+12x_0^2r_0^4+12y_0^2r_0^4+4r_0^6)/D=0
\nonumber \\
\ee
with the denominator

\be
D\propto x_0(r_0^2+(x_0+1)^2)(3x_0^2+x_0^3+r_0^2+3xr^2)
\ee
The positivity of $r_0^2$ defines a 3-dimensional surface $r_0(x,0, y_0)$. In Fig.~\ref{SF-NEGATIVEx} (front surface)
we show a cut of the surface through the plane $x_0=0$ for $x_0<0$, while in Fig.~\ref{SF-POSITIVEx} (back surface)
we show  a cut of the same surface for $x_0>0$. As shown, the front surface is composed of an inner and outer surfaces. 
In both figures, the side is $y_0$, the height is $r_0$ and the depth is $x_0$.
The inner surface excludes a region in parameter space $(r_0, x_0, y_0)$ where no mixed condensate develops. 
Indeed, for $D\neq 0$ and  in the plane $r_0=0$ the surface defines a curve

\begin{eqnarray}
\label{PLANAR}
&&(1 + 2x_0 + x_0^2 + y_0^2)\nonumber \\
&&(x_0-3x_0^2 + 2x_0^4 - 3y_0^2 + 4x_0^2y_0^2+ 2y_0^4)=0 
\end{eqnarray}
The second contribution in (\ref{PLANAR}) is the boundary curve (\ref{BOUND2}). 
 It contains two connected pieces also, the inner part is just the intersection between the small island and the (x,y) plane.
 The first contribution in (\ref{PLANAR}) vanishes at the point   $x_0=-1,y_0=r_0=0$ which is where the back surface
 is pinching the plane $r_0=0$ in~Fig.~\ref{SF-NEGATIVEx}. This zero is not spurious as can be seen through the plane
 $(x_0,r_0)$ plane, where (\ref{LONGG}) simplifies to
 
\begin{eqnarray}
&&(-1+x_0^2+r_0^2)\\&&(-x_0+x_0^2+4x_0^3+2x_0^4+r_0^2+4x_0r_0^2+4x_0^2r_0^2+2r_0^4)=0\nonumber 
\end{eqnarray}
There is a continuous limit to $x_0=-1$ from the parameter space for a mixed condensate for arbitrary small $r_0$. The
factor $(r_0^2+x_0^2-1)$ cannot be cancelled by the denominator $D$ which is non-vanishing at this point. 
Thus, the intersection of the condensation region with the real axis  shows $x_0=-1$ as an accumulation point.
Now, within the real axis and the branch of the surface determined by $r_0^2=1-x_0^2$, which is 
the "outer layer" of the condensation region, the first equation for (\ref{SADDLEX})  can be solved at once

\be
x=\frac{1}{4}(7+6x_0)\qquad\qquad y=0
\ee
The point $x_0=-1,y_0=0,r_0=0$ correspond to the chiral point $z=\mu^2=\frac{1}{4}$. But, at  
this point, the saddle point momentum and the free energy   $f$ in (\ref{fP}) diverge.
 The chiral point $z=\mu^2$ is a singular point  on the surface $r_0=r_0(x_0,y_0)$  defined by (\ref{LONGG}), as the outer surface develops two sharp holes that connects through a pinch. The standard $1/N$ expansion breaks down.

\subsection{Chiral microscopic universality}

To analyze the chiral point more accurately we need an alternative to the the standard $1/N$ expansion,
that resumes a class of corrections around the chiral point. 
For that it is useful to go back to (\ref{NEWx}-\ref{NEWxxx}) and use the following microscopic
rescaling at the chiral point

\begin{eqnarray}
\label{XSCAX}
&&z-\mu^2\rightarrow \frac{z}{N^2},\mu^2\rightarrow \frac{\mu^2}{N}\nonumber\\
&&w\rightarrow \frac{w}{N^2},p\rightarrow Np, k\rightarrow Nk
\end{eqnarray}
In leading order (\ref{NEWx}) simplifies

\begin{eqnarray}
&&\Psi(\tau, \mu^2/N+z/N^2, 0)\approx \nonumber\\
&& \int \frac{d^2 p kdkd^2 z^\prime d^2w^\prime}{(2\pi)^3}\frac{e^{F}}{(p^2+k^2)^2}
\end{eqnarray}
with

\begin{eqnarray}
&&F\approx +\frac{N(\bar z^\prime +z^\prime)}{\tau}+N\,{\rm ln}(|z^\prime|^2+(r^\prime)^2)\nonumber\\
&&+\frac{p z+\bar p \bar z}{2}+kr+2\frac{\bar p z^\prime +p \bar z^\prime}{\tau^2(\bar p p+k^2)}-\frac{4\mu^2\bar p p}{\tau(k^2+\bar p p)}\end{eqnarray}

We use the saddle point solution in $z^\prime, r^\prime$ to undo this double integration, i.e. 
$z^\prime =-1+\alpha$ and $r^\prime=0+\beta$. The resulting Gaussian integrations in $\alpha, \beta$ 
decouple. Thus

\begin{eqnarray}
\label{X01}
&&\Psi(\tau, \mu^2/N+z/N^2, 0)\nonumber\\
&&\approx \int \frac{d^2p \,kdk}{(2\pi)^2}
\frac{e^{\frac{p z+\bar p \bar z}{2}-2\frac{\bar p +p }{\tau(\bar p p+k^2)}-\frac{4\mu^2\bar p p}{(k^2+\bar p p)\tau}}}{(p^2+k^2)^2}
\end{eqnarray}
The $k$-integration in (\ref{X01}) can be done by expanding part of the exponent,

\begin{eqnarray}
\label{SHORT}
&&\Psi(\tau, \mu^2/N+z/N^2, 0)\nonumber\\
&&\approx \int \frac{dp d\bar p}{(2\pi)^2}\sum_{n=0}^{\infty}\int kdk
\frac{(-2(p+\bar p)-4\mu^2\bar p p)^n}{\tau^n(k^2+\bar p p)^{n+2}n!}e^{\frac{\bar p \bar z+pz}{2}}\nonumber\\
&&=\int \frac{dp d\bar p}{(2\pi)^2}\sum_{n=0}^{\infty}\frac{1}{\bar p p}
\frac{(-2(p+\bar p)-4\mu^2\bar p p)^n}{\tau^n(n+1)!(\bar p p)^{n}}e^{\frac{\bar p \bar z+pz}{2}}\nonumber\\
&&=\int_{0}^{1}dt\int\frac{dp d\bar p}{(2\pi)^2}\frac{1}{\bar p p}
e^{-t\frac{2(\bar p+p+4\mu^2\bar p p)}{\bar p p\tau }}e^{\frac{\bar p \bar z+pz}{2}}\nonumber \\
&&=\int^{1}_0dte^{-\frac{4\mu^2t}{\tau}}\int\frac{dp}{2\pi p}e^{pz/2-2t/p\tau}\int\frac{d\bar p}{2\pi \bar p}e^{\bar p\bar z/2-2t/\bar p \tau}\nonumber\\
\end{eqnarray}
We note that

\be
\xi (z)=\int\frac{dp}{2\pi p}e^{pz/2-2t/p\tau}
\ee
satisfies

\be
\left(z\partial_z^2+\partial_z+\frac t\tau \right)\xi (z)=0
\ee
$\xi(z)=J_0(2\sqrt{tz/\tau})$ is the unique solution regular at the origin. 
So as long as we can set the $p$-integration contour in (\ref{SHORT}) 
so that ${\Psi}$ is regular at $z=0$, we have $\xi(z)=J_0(2\sqrt{tz/\tau})$.
Thus the microscopic form of the characteristic determinant at the chiral point is

\begin{eqnarray}
\label{X05}
&&\Psi(\tau, \mu^2/N+z/N^2, 0)\approx \nonumber\\
&&\approx \tau^{2N} \int_{0}^{1}dt\,e^{-\frac{4\mu^2t}{\tau}}
\left|J_0\left({2\sqrt{\frac {tz}{\tau}}}\right)\right|^2
\end{eqnarray}
Here  $\mu^2$ is short for the fixed combination $N\mu^2$ at large $N$, as defined through the re-scaling in  (\ref{XSCAX}).
This is the fourth main result of this paper.

 In contrast to (\ref{CHAMU}) the scaling law for the characteristic determinant
at the chiral point (\ref{X05}) does not record the quenched chiral condensate at finite $\mu$. We note the similarity of 
(\ref{X05}) with the microscopic law at the chiral point for the phase phase quenched density of eigenvalues in~\cite{INT}.

For completeness, we note that (\ref{CLOSED}) simplifies at the chiral point for large $N$, 
but fixed $N\mu^2$ and $N^2z$, i.e. 
$\mu^2\rightarrow \mu^2/N$, $z\rightarrow \mu^2/N+z/N^2$ and $w\rightarrow w/N^{2}$.
In the microscopic limit, the resulting differential equation is

\begin{eqnarray}
\label{CLOSEDx}
&&\Bigg(2\,(\partial_z+\partial_{\bar z})+(z\partial_z^2
+\bar z\partial_{\bar z}^2) \Bigg.\nonumber \\
&&-(z+\bar z)\partial^2_{\bar ww}\Bigg. +(\partial_z+\partial_{\bar z})(w\partial_w+\bar w\partial_{\bar w})
+\nonumber \\ &&\frac{4\mu^2}{\tau}\frac{(\partial_z+\partial_{\bar z})\,\partial^2_{\bar w w}}{(\partial_z\partial_{\bar z}+\partial^2_{\bar w w})}\Bigg)\bf\Psi\approx 0
\end{eqnarray}
The solution to (\ref{CLOSEDx}) as $w\rightarrow 0$ is (\ref{X05}).

\section{Conclusions}

The QCD Dirac spectrum at finite chemical potential contains subtle information on the chiral
dynamics of light quarks in matter. Using a random matrix model, we have shown that the characteristic determinant of 
phase quenched QCD with $N_f=4$  massless quarks is related to the characteristic 
determinant of a class of deformed Wishart matrices through a conformal mapping in the
space of eigenvalues.

We have constructed a stochastic evolution for the deformed Wishart matrices by allowing the Gaussian
weights in random matrices to diffuse. The mathematical diffusion time is identified with the stochastic
time. We have derived an exact solution to the stochastic diffusion equation for any finite $N$ and
derived the explicit time-evolving boundary condition of the envelope of the deformed Wishart eigenvalues
through a semi-classical expansion. 

Contrary to the lore of random matrix approaches to QCD~\cite{VERBAARSCHOTREVIEW}, which focus on the spectral density  distributions of the Dirac operator, we studied here the characteristic determinant.  Following a recent observation
in~\cite{NOWAK}, that the spectral evolution of non-Hermitean and non-normal ensembles involves a hidden complex variable $w$~\cite{BLUE}, we have embedded the Dirac operator into this general algebraic structure.  The explicit dependence on $w$  is key to closing and  obtaining  the main evolution equations   (\ref{CLOSED}-\ref{WW2}).  While the complex variable  $z$ reflects on the  standard quark condensate, the complex variable  $w$ reflects on the "spurious" quark condensate~\cite{STEPHANOV, US,BLUE},  whose formation is an artifact of quenching or ignoring the phase of the fermionic determinant.  This evolution equation is exact for any finite $N$.  This fact allows to trace the co-evolution of both type of condensates, and  to perform all kinds of rescalings in the vicinity of the physically pertinent points.

In general, the complex eigenvalues form a droplet in the complex eigenvalue plane that 
breaks conformal symmetry. The deformed Wishart droplet deforms but never breaks. Its boundary is sharp at large $N$, 
and smoothens out in $1/N$  through a universal edge function in leading order.  At the chiral
point, the characteristic determinant follows universally from a pertinent Bessel kernel.

The edge universality and the chiral universality
derived in this work can be numerically checked. In particular, the microscopic scaling law
at the edge of the spectrum scales with the  chiral condensate at finite $\mu$,
allowing for its possible extraction directly  from  current and quenched lattice data. 
In practice and in the absence of an apparent edge in real QCD spectra,
 this can be achieved by rescaling the numerical eigenvalues within a sliding window along the real axis,
and checking for the edge scaling law for the Dirac spectrum using for instance 
(\ref{EDGREALDIRAC}-\ref{DENSMICRO}), or along the $w$-axis using (\ref{ANXXX}-\ref{ANXXXX}).
When extended to finite temperature, this practical analysis  will allow for a determination of the QCD 
phase diagram solely from the quenched lattice simulations,
without having to solve the QCD sign problem, a major achievement in this field.

Finally and on general grounds, most of our results for the deformed
Wishart ensemble may be of relevance to a wider audience of practitioners using non-hermitean random matrix 
methods in  the fields of wireless communication, biological and neural information,   and finance~\cite{NONHERM}. 

\section{\ Acknowledgements}

We thank Piotr Warcho\l{} and Jacek Grela for a discussion. 
We would like to thank the organizers 
of  the Workshop  on Random Matrix Theory, Integrable Systems, and Topology in Physics organized at the Simons Center for Geometry and Physics at Stony Brook, where this work was initiated.
This work  is  supported in part  by the U.S. Department of Energy under Contracts No.
DE-FG-88ER40388 (YL and IZ) and by the Grant DEC-2011/02/A/ST1/00119 of the National Center of Science  and 
by the grant from the Simons Foundation (MAN).

\section{Appendix I}

It is instructive to re-check the edge scaling laws (\ref{SCA1}-\ref{EDGEX}) near $x_0/\tau=1$ 
directly from (\ref{CLOSED}). Recall that  this point maps onto the outer edge $z(\tau)=4\tau-3\mu^2$ in the
Wishart ensemble and ${\bf z}(\tau)=\pm 2\sqrt{\tau-\mu^2}$ in the Dirac ensemble. 
At $x_0=\tau$, the leading contribution to the characteristic determinant in large $N$ is

\be
\label{F0}
f_0=\frac{-2\mu^2}{\tau}+2\,{\rm ln}\tau
\ee
Inserting the re-scaling 

\be
\label{SCAXX}
z(\tau) =4\tau-3\mu^2+\frac{\eta}{{N}^{\frac 12}},\qquad w=\frac{\omega}{N^{\frac 34}}
\ee
and the re-scaled ansatz

\begin{eqnarray}
\label{ANXX}
&&\Psi(\tau, 4\tau-3\mu^2+\eta/{N}^{\frac 12}, \omega/N^{\frac 34})\nonumber\\
&&\approx \tau^{2N}e^{-\frac{2N\mu^2}{\tau}}e^{\frac{\sqrt{N}}{2\tau}(\eta+\bar \eta)}
\varphi (\tau, \eta,\bar \eta, \omega)
\end{eqnarray}
in  (\ref{SCA1}-\ref{EDGEX}),  yield the identity to order $N^2$

\be
\left(\frac{2\mu^2}{\tau^2}-\frac{2}{\tau}\right)\varphi =\left(\frac{2\mu^2}{\tau^2}-\frac{2}{\tau}\right)\varphi 
\ee
which is satisfied whatever $\varphi$. At order $N^{\frac 32}$,  $\varphi$ is fixed by

\be
\label{NXX1}
(8\tau-8\mu^2)\partial_{\bar \omega \omega}^2\varphi
+\frac{4\mu^2}{\tau}(\partial_{\eta}+\partial_{\bar \eta})\varphi+\frac{1}{4\tau^2}(\bar \eta +\eta)\varphi=0\nonumber \\
\ee
The formal  solution to (\ref{NXX1}) is

\begin{eqnarray}
\label{ONEX}
&&\varphi(\tau, \eta,\bar \eta, \omega)\approx \\
&&\int_0^\infty d\lambda\, \kappa (\lambda)\,I_0\left(\frac{\sqrt{\lambda}|\omega|}{\sqrt{2(\tau-\mu^2)}}\right)
\, e^{-\frac{\tau}{4\mu^2}(\frac{\eta+\bar \eta}{4\tau}+\lambda)^2}g(\tau, \eta-\bar \eta)\nonumber
\end{eqnarray}
with arbitrary positive weight $\kappa$ and arbitrary $g$. 
A comparison of (\ref{ONEX}) to (\ref{COMPX}) fixes $\kappa$ and $g$ to be respectively

\begin{eqnarray}
&&\kappa (\lambda)\approx \frac{\tau}{(\tau-\mu^2)}\nonumber\\
&&g(\tau, \eta-\bar\eta)\approx e^{-\frac{(\eta-\bar\eta)^2}{64\tau\mu^2}}
\end{eqnarray}
Thus

\begin{eqnarray}
\label{ANXXX}
&&\Psi(\tau, 4\tau-3\mu^2+\eta/{N}^{\frac 12}, \omega/N^{\frac 34})\nonumber\\
&&\approx \tau^{2N+1}
\frac{e^{-\frac{2N\mu^2}{\tau}}e^{\frac{\sqrt{N}}{2\tau}(\eta+\bar \eta)}}{(\tau-\mu^2)}\\
&&\int_0^\infty d\lambda\,I_0\left(\frac{\sqrt{\lambda}|\omega|}{\sqrt{2(\tau-\mu^2)}}\right)
\, e^{-\frac{\tau}{4\mu^2}(\frac{\eta+\bar \eta}{4\tau}+\lambda)^2-\frac{(\eta-\bar\eta)^2}{64\tau\mu^2}}\nonumber
\end{eqnarray}
(\ref{ANXXX}) shows that along the $\omega$-direction, the characteristic determinant at the real 
edge of the Wishart and therefore the Dirac spectrum through (\ref{TRANX}) for $x_0=\tau$, scales with the 
chiral condensate (\ref{QQX}). Thus the edge scaling law (\ref{ANXXX})  allows for a possible measurement
of the chiral condensate at the edge of the complex eigenvalue droplet along the $w$-direction,

\begin{eqnarray}
\label{ANXXXX}
&&\Psi_D(\tau, {\bf z}_0, \omega/N^{\frac 34})\\
&&\approx \tau^{2N+1}
\frac{e^{-\frac{2N\mu^2}{\tau}}}{(\tau-\mu^2)}
\int_0^\infty d\lambda\,I_0\left(\frac{\sqrt{\lambda}|\omega|}{\sqrt{2(\tau-\mu^2)}}\right)
\, e^{-\frac{\tau\lambda^2}{4\mu^2}}\nonumber
\end{eqnarray}
(\ref{ANXXXX}) is the analogue of (\ref{EDGREALDIRAC}) along the 
$w$-direction.

\section{Appendix II}

In this Appendix we will show that the 
characteristic determinant for the 2-matrix model~\cite{OSB,AKE} obeys a closed form equation
analogous to (\ref{CLOSED}) and share the same microscopic universality near the chiral Dirac point. For that, we define

\be
\label{X21}
{\bf\Phi}(\tau, z, w)=\left<{\rm det} \left((z-{\mathbb W})(\bar z-{\bf  \bar {\mathbb W}})+|w|^2\right)\right>
\ee
with the newly deformed  and non-hermitean Wishart-like matrix 

\be
{\mathbb W} =({\bf A}^\dagger-i\mu {\bf B}^\dagger)({\bf A}-i\mu {\bf B})
\ee
The relationship between the 2-matrix Dirac spectrum at finite $\mu$ with eigenvalues ${\bf z}$ and the 
newly deformed Wishart ${\mathbb W}$ spectrum with eigenvalues $z$ is through the new mapping $z={\bf z}^2$.
The averaging in (\ref{X21}) is now carried using the double Gaussian weight

\be
\label{XM2}
{\bf P}(\tau, {\bf A, B})\approx e^{-\frac N\tau {\rm Tr} ({\bf A}^\dagger{\bf A}+{\bf B}^\dagger{\bf B})   }
\ee
Using the same arguments as those developed above, we unwind $\partial_\tau {\bf \Phi}$ in terms 
of Grassmannians, undo the ${\bf A, B}$ integrations and carry the Grassmannian integrations
by trading them with partial derivatives in $(z, \bar z, w, \bar w)$. The result is a closed form equation

\begin{eqnarray}
\label{WW2}
&&N\partial_{\tau}{\bf \Phi}= -2(\partial_z+\partial_{\bar z}){\bf \Phi}\nonumber\\
&&+\left(\frac{\mu^2}\tau-1\right)\nonumber\\
&&\times
\left(z\partial_z^2+\bar z\partial_{\bar z}^2
+\frac{1}{2}\left(\partial_z+\partial_{\bar z})(w\partial_w+\bar w\partial_{\bar w}\right)\right){\bf \Phi}\nonumber \\
&&+\left(\left(1+\frac{\mu^2}\tau\right)\right.\nonumber\\
&&\times \left. \left((z+\bar z)\partial^2_{\bar w w}-
\frac 12 (\partial_z+\partial_{\bar z})(w\partial_w+\bar w\partial_{\bar w})\right)\right){\bf \Phi}\nonumber\\
&&+\left({\left(1-\frac{\tau}{N}(\partial_z+\partial_{\bar z})\right)+\frac{4\mu^2}\tau\left(1+\frac {\mu^2}\tau\right)\partial^2_{\bar w w}}\right)\nonumber\\
&&\times\left({\left|1+\left(\frac{\mu^2}\tau -1\right)\frac{\tau}{N}\partial_z\right|^2
+\left|\left(1+\frac{\mu^2}\tau\right)\frac{\tau}{N}\partial_{w}\right|^2}
\right)^{-1}{\bf \Phi}\nonumber\\
\end{eqnarray}
which is the analogue of (\ref{CLOSED}) for the 2-matrix model. 
Here, the chiral point in the Dirac spectrum or ${\bf z}=0$ maps onto the Wishart point $z=0$. 
Using a similar microscopic re-scaling or $z\rightarrow\frac{z}{N^2},w\rightarrow \frac{w}{N^{2}}$ 
around the chiral point in (\ref{WW2}) yields (\ref{CLOSEDx}). The characteristic determinant for
both the 1-matrix and 2-matrix  models shares the same microscopic universality at the chiral point.
This observation extends to the characteristic determinant the universality noted at the chiral point for the 
microscopic density for both the 1- and 2-matrix models~\cite{INT}.

 \vfill
\end{document}